\documentclass[openacc]{rstransa}

\newtheorem{theorem}{\bf Theorem}[section]

\newtheorem{definition}{\bf Definition}[section]

\graphicspath{{./SERivlin-figs/}}
\begin{document}

\title{Likely equilibria of the stochastic Rivlin cube}

\author{
L. Angela Mihai$^{1}$, Thomas E. Woolley$^{1}$ and Alain Goriely$^{2}$}

\address{$^{1}$School of Mathematics, Cardiff University, Senghennydd Road, Cardiff, CF24 4AG, UK\\
$^{2}$Mathematical Institute, University of Oxford, Woodstock Road, Oxford, OX2 6GG, UK}

\subject{applied mathematics, mathematical modelling, mechanics}


\corres{L. Angela Mihai\\
\email{MihaiLA@cardiff.ac.uk}}

\begin{abstract}
The problem of the Rivlin cube is to determine the stability of all homogeneous equilibria of an isotropic incompressible hyperelastic body under equitriaxial dead loads. Here, we consider the stochastic version of this problem where the elastic parameters are random variables following standard probability laws. Uncertainties in these parameters may arise, for example, from inherent data variation between different batches of homogeneous samples, or from different experimental tests. As for the deterministic elastic problem, we consider the following questions: what are the likely equilibria and how does their stability depend on the material constitutive law? In addition, for the stochastic model, the problem is to derive the probability distribution of deformations, given the variability of the parameters.
\end{abstract}


\begin{fmtext}
\section{Introduction}\label{sec:intro}
	
The so-called ``Rivlin cube'' is a classic problem of mechanics  that has played a central role in the development of new concepts in nonlinear elasticity \cite{James:1985}. The problem, first introduced by Rivlin (1948) \cite{Rivlin:1948:II}, is to obtain all equilibria of a homogeneous isotropic incompressible hyperelastic cube on which equal triaxial dead loads are applied to each face. If we only consider homogeneous triaxial stretches, the natural questions are: \emph{what are the possible equilibrium states and how does their stability depend on the material constitutive law?} The problem then becomes a tractable case of bifurcation analysis for the algebraic equations relating stretches and dead-load tractions \cite{Hill:1957,Beatty:1967,Rivlin:1974,Ball:1983:BS}. The result can be summarised as a two-dimensional bifurcation diagram for the  stretch against dead loads, depicting the possible equilibrium states and their stability.
\end{fmtext}


\maketitle

\begin{figure}[htbp]
	\begin{center}
		\includegraphics[width=0.99\textwidth]{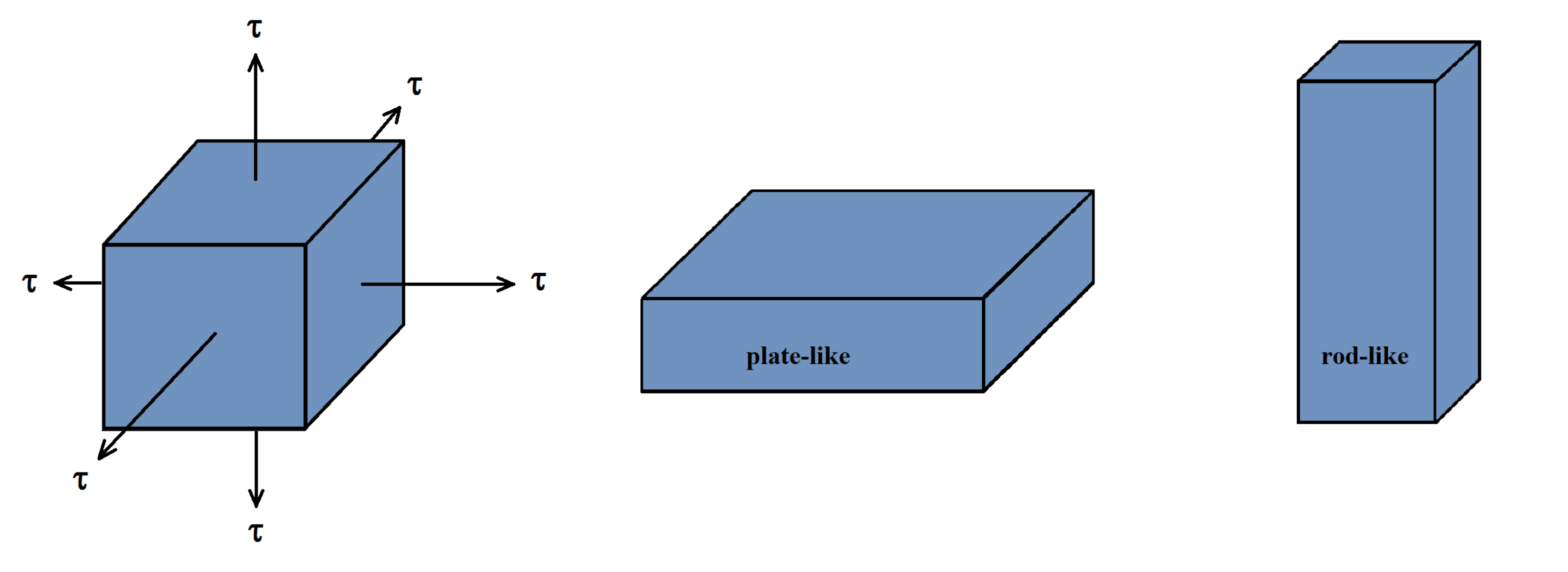}
		\caption{Schematic of Rivlin's cube under normal dead-load tractions, $\tau>0$, uniformly distributed on all faces in the reference configuration, showing the reference state and the plate-like or rod-like deformations.}\label{rivlincube}
	\end{center}
\end{figure}

We recall that, for a dead load, the magnitude and direction of the force are maintained independently of how the body deforms, and that homogeneous deformations are the same regardless of the geometry of the body. Furthermore, the homogeneous deformations are universal in the sense that they can be maintained in every homogeneous isotropic hyperelastic body by application of suitable dead-load tractions  \cite{Ericksen:1954,Ericksen:1955,Singh:1965:SP,Shield:1971,Yavari:2016:anelastic}. From these definitions, it can be shown that, under equitriaxial surface dead loads, a homogeneous deformation is equivalent to a homogeneous triaxial stretch \cite{Ball:1983:BS}. For a cube of neo-Hookean material, Rivlin found that (neutrally) stable non-trivial triaxial deformations with two equal stretches are possible if the equitriaxial tractions are sufficiently large \cite{Rivlin:1948:II,Hill:1957,Beatty:1967,Rivlin:1974}. In addition to these so-called ``plate-like'' and ``rod-like'' equilibrium states (figure~\ref{rivlincube}), for a similar cube of Mooney-Rivlin material, it was shown in \cite{Ball:1983:BS} that neutrally stable equilibria with three unequal stretches are also possible.

Extensions to the case of a cube of compressible Mooney-type material were studied in \cite{Tarantino:2008}, while the case of anisotropic materials was analysed in \cite{Soldatos:2006}. For an isotropic cube, when the three pairs of equal and opposite forces differ from each other by a small amount \cite{Rivlin:1948:II}, or if two pairs are the same but different from the third by a small amount \cite{Sawyers:1976}, neutrally stable homogeneous triaxial deformations with three different stretches can occur as the reference state becomes unstable. For further discussions on stability analysis under small perturbations, we refer to \cite{Golubitsky:1979:GS}. However, the cases involving compressible or anisotropic materials, or perturbed loading will not be treated here.

In this study, we consider the Rivlin cube in the context of stochastic elasticity, where the elastic parameters appearing in the strain-energy function are random variables satisfying standard probability laws \cite{Staber:2015:SG,Staber:2016:SG,Staber:2017:SG,Staber:2018:SG,Mihai:2018a:MWG}.  As this framework takes into account both the average values of the elastic parameters, as well as their variability, our choice reflects the inherent variation in material properties or in testing protocols \cite{Farmer:2017,Hughes:2010:HH,Oden:2018}. Specifically, stochastic representations of isotropic incompressible hyperelastic materials were proposed in \cite{Staber:2015:SG}, then Ogden-type stochastic strain-energy functions were calibrated to experimental data for soft tissues under uniaxial loads in \cite{Staber:2017:SG}. Compressible versions of these stochastic models were presented in \cite{Staber:2016:SG}, while anisotropic stochastic models were calibrated to vascular tissue data in \cite{Staber:2018:SG}. The general strategy relies on the maximum entropy principle for a discrete probability distribution introduced by Jaynes (1957) \cite{Jaynes:1957a,Jaynes:1957b,Jaynes:2003} and based on the notion of  entropy (or uncertainty) defined by Shannon (1948) \cite{Shannon:1948,Soni:2017:SG}. In \cite{Mihai:2018a:MWG}, we extend the stochastic modelling approach of \cite{Staber:2017:SG} to various hyperelastic models and to multiaxial deformations as well, and deploy Bayes' theorem \cite{Bayes:1763} to select a model among competing models calibrated to experimental data.

Stochastic elastic modelling raises a number of important questions about the behaviour of materials. The first one, considered here, is the problem of stability. As a first step, we consider the stochastic Rivlin cube and  extend the classic results to the case where the properties of the chosen cubes, although homogeneous, are no longer assured, but rather sampled from a specified probability density function. The problem under consideration can be interpreted in the following way: Imagine a population (or set) of cubes, where each cube is made from a single homogeneous isotropic incompressible hyperelastic material, with the elastic parameters not known with certainty, but distributed according to a known probability function. Then, for every homogeneous isotropic hyperelastic cube, the finite elasticity theory applies. Assuming that each cube is subject solely to equitriaxial dead-load tractions, uniformly distributed on all faces in the reference configuration, and that every cube deforms by homogeneous triaxial stretch, the problem is to find the probability distribution of the stable deformations of a randomly chosen cube. Note that, as homogeneous deformations are independent of the geometry, assuming uncertainty in the shape of the body will not change the problem, so the unit cube is chosen for definiteness. For the deterministic elastic problem, we find that stable non-trivial triaxial deformations are possible if the tractions are sufficiently large. Moreover, in the stochastic case, the probabilistic nature of the solution reflects the variability in the structural properties.

We begin, in section~\ref{sec:problem}, by posing the Rivlin cube problem in the deterministic elastic context, and also in the new, stochastic elastic setting. In section~\ref{sec:estable}, we review the solution to the elastic problem of a cube of neo-Hookean or Mooney-Rivlin material as particular cases of the general stability analysis developed in \cite{Ball:1983:BS}. This general analysis further permits us to discuss, for the first time, the case of a Mooney-Rivlin material where one of the model coefficients may take negative values, while the shear modulus under small strain is always positive. For completeness, and to help to see this, we provide in appendix~\ref{sec:append} the proof of the relevant result from \cite{Ball:1983:BS}. General definitions for various linear and nonlinear elastic moduli and their mutual universal relations in finite elasticity can be found, for example, in the review article \cite{Mihai:2017:MG}. For the stochastic version of the Rivlin cube problem, in section~\ref{sec:sestable}, we derive the probability distribution of stable equilibria from the stochastic model coefficients, and illustrate our results through novel stochastic bifurcation diagrams. As shown by these new diagrams, in this case, one cannot simply talk about \emph{'equilibria'}, but \emph{'likely equilibria'} observed at a given applied load with a given probability. Concluding remarks and a further outlook are included in section~\ref{sec:conclude}.

\section{Problem formulation}\label{sec:problem}

In this section, first, we formulate the static equilibrium problem of a cube of incompressible neo-Hookean or Mooney-Rivlin material, subject to equitriaxial dead-load tractions and deforming by a homogeneous triaxial stretch. We then recast this problem in the case of stochastic materials. The stability of the equilibrium solutions to these problems are discussed in the next sections. 

Here, a stochastic hyperelastic material represents an ensemble of hyperelastic material models that are similar in form, and for which every model parameter is a random variable that lies between known bounds and satisfies standard probability laws. A model parameter, $C$, is described in terms of the \emph{mean value}, $\underline{C}$, and the \emph{variance}, $\text{Var}[C]$, which contains information about the range of values about the mean value. The corresponding \emph{standard deviation} is $\|C\|=\sqrt{\text{Var}[C]}$. Our approach combines mathematics of information and finite elasticity, and relies on the following key assumptions \cite{Mihai:2018a:MWG}:

	\noindent\textbf{(A1) Material objectivity:} The principle of material objectivity (frame indifference) states that constitutive equations must be invariant under changes of frame of reference. It requires that the scalar strain-energy function, $W=W(\textbf{F})$, depending only on the deformation gradient $\textbf{F}$, with respect to the reference configuration, is unaffected by a superimposed rigid-body transformation (which involves a change of position) after deformation, i.e., $W(\textbf{R}^{T}\textbf{F})=W(\textbf{F})$, where $\textbf{R}\in SO(3)$ is a proper orthogonal tensor (rotation). Material objectivity is guaranteed by considering strain-energy functions defined in terms of invariants.
	
	\noindent\textbf{(A2) Material isotropy:} The principle of isotropy requires that the strain-energy function is unaffected by a superimposed rigid-body transformation prior to deformation, i.e., $W(\textbf{F}\textbf{Q})=W(\textbf{F})$, where $\textbf{Q}\in SO(3)$. For isotropic materials, the  strain-energy  function is a symmetric function of the  principal stretches, $\{\lambda_{i}\}_{i=1,2,3}$, i.e.,  $W(\textbf{F})=\mathcal{W}(\lambda_{1},\lambda_{2},\lambda_{3})$.
	
	\noindent\textbf{(A3)] Baker-Ericksen inequalities:} In addition to the fundamental principles of objectivity and material symmetry, in order for the behaviour of a hyperelastic material to be physically realistic, there are some universally accepted constraints on the constitutive equations. Specifically, for a hyperelastic body, the Baker-Ericksen  (BE) inequalities  state that  \emph{the greater principal Cauchy stress occurs in the direction of the greater principal stretch} \cite{BakerEricksen:1954}, i.e.,
	\begin{equation}\label{eq:BE}
	\left({T}_{i}-{T}_{j}\right)\left(\lambda_{i}-\lambda_{j}\right)>0\quad \mbox{if}\quad \lambda_{i}\neq\lambda_{j},\quad i,j=1,2,3,
	\end{equation}
	where $\{\lambda_{i}\}_{i=1,2,3}$ and $\{T_{i}\}_{i=1,2,3}$ denote the principal stretches and the principal Cauchy stresses, respectively, and the strict inequality ``$>$'' is replaced by ``$\geq$'' if any two principal stretches are equal \cite{BakerEricksen:1954}. 
	In particular, under uniaxial tension, the deformation is a simple extension in the direction of the tensile force if and only if the BE inequalities hold \cite{Marzano:1983}. Under these mechanical constraints, the shear modulus of the material, under finite strains, is positive \cite{Mihai:2017:MG}. 
	
	\noindent\textbf{(A4) Finite mean and variance for the random shear modulus:} We assume that for any given finite deformation, the random shear modulus, $\mu$, and its inverse, $1/\mu$, are second-order random variables, i.e., they have finite mean value and finite variance \cite{Staber:2015:SG,Staber:2016:SG,Staber:2017:SG}.
	
While (A1)-(A3) are well known requirements in isotropic finite elasticity \cite{goriely17,Ogden:1997,TruesdellNoll:2004}, (A4) describes physically realistic expectations on the random shear modulus, to be characterised by a probability distribution.

\subsection{Elastic setting}\label{sec:elastic}
We consider first an elastic cube, occupying the domain $V=(0,1)\times(0,1)\times(0,1)\subset\mathbb R^{3}$ in the reference configuration, and made from a homogeneous isotropic incompressible Mooney-Rivlin material characterised by the usual  strain-energy function \cite{Mooney:1940,Rivlin:1948:IV}
\begin{equation}\label{eq:mr}
\mathcal{W}_{mr}(\lambda_{1},\lambda_{2},\lambda_{3})=\frac{\mu_{1}}{2}\left(\lambda_{1}^2+\lambda_{2}^2+\lambda_{3}^2-3\right)
+\frac{\mu_{2}}{2}\left(\frac{1}{\lambda_{1}^2}+\frac{1}{\lambda_{2}^2}+\frac{1}{\lambda_{3}^2}-3\right),
\end{equation}
where $\mu_{1}$ and $\mu_{2}$ are given constant coefficients, and $\mu=\mu_{1}+\mu_{2}>0$ is the shear modulus under small strain \cite{Mihai:2017:MG}. We assume that this cube is subject to constant normal tension or compression, $\tau$, that is uniformly distributed on all faces in the reference configuration, in the absence of body forces. Assuming that the resulting deformation is a homogeneous triaxial stretch, the deformation gradient is constant, and the gradient tensor $\textbf{F}$, measured from a reference configuration, is
\begin{equation}\label{eq:triaxial}
\textbf{F}=\mathrm{diag}(\lambda_{1},\lambda_{2},\lambda_{3}),
\end{equation}
with the constants $\lambda_{i}>0$, $i=1,2,3$, satisfying the incompressibility condition,
\begin{equation}\label{eq:inc}
\det\textbf{F}=\lambda_{1}\lambda_{2}\lambda_{3}=1.
\end{equation}
In this case, the Cauchy equations for equilibrium are automatically satisfied \cite{TruesdellNoll:2004,Ogden:1997,goriely17}, and the deformation is fully specified by the values of the stresses at the boundary. Therefore, possible deformations are  given by the solutions of an algebraic system of equations as explained next. Under the deformation \eqref{eq:triaxial}, the principal components of the Cauchy stress tensor $\mathbf{T}(\textbf{x})$, with $\textbf{x}=(x_{1},x_{2},x_{3})^{T}$ the Cartesian coordinates in the current  configuration, are
\begin{equation}\label{eq:sigmai:mr}
{T}_{i}=-p+\lambda_{i}\frac{\partial\mathcal{W}_{mr}}{\partial\lambda_{i}}
=-p+\mu_{1}\lambda_{i}^2-\frac{\mu_{2}}{\lambda_{i}^2}, \qquad i=1,2,3,
\end{equation}
where $p$ is the Lagrange multiplier for the incompressibility constraint \eqref{eq:inc}. The corresponding first Piola-Kirchhoff stress tensor $\textbf{P}(\textbf{X})=J\mathbf{TF}^{-T}$, with $\textbf{X}=(X_{1},X_{2},X_{3})^{T}$ the Cartesian coordinates in the reference configuration, has the principal components
\begin{equation}\label{eq:pki:mr}
P_{i}=\frac{T_{i}}{\lambda_{i}}\\
=-\frac{p}{\lambda_{i}}+\mu_{1}\lambda_{i}-\frac{\mu_{2}}{\lambda_{i}^3}, \qquad i=1,2,3.
\end{equation}
In the absence of body forces, the elastic equilibrium equations in the reference configuration are
\begin{equation}\label{eq:equilibrium:1}
\frac{\partial P_{i}}{\partial X_{i}}=0,\qquad i=1,2,3.
\end{equation}
Equivalently, in the current configuration,
\begin{equation}\label{eq:equilibrium:2}
\frac{\partial{T}_{i}}{\partial x_{i}}=0,\qquad i=1,2,3.
\end{equation}
Thus, by \eqref{eq:sigmai:mr}, the Lagrange multiplier $p$ satisfies
\begin{equation}\label{eq:p}
\frac{\partial p}{\partial x_{i}}=0,\qquad i=1,2,3,
\end{equation}
i.e., $p$ is constant. The equitriaxial surface tractions, $\tau$, are constant in the reference configuration (dead loading), i.e.,
\begin{equation}\label{eq:tractions:mr}
P_{i}=\tau, \qquad i=1,2,3,
\end{equation}
or equivalently, by \eqref{eq:pki:mr},
\begin{equation}\label{eq:quartic}
\mu_{1}\lambda_{i}^2-\frac{\mu_{2}}{\lambda_{i}^2}-\tau\lambda_{i}-p=0, \qquad i=1,2,3.
\end{equation}
Given the dead load $\tau$, the problem is to \emph{find all possible homogeneous triaxial deformations} and to \emph{identify the stable equilibrium states}. The equilibria and stability of an elastic cube of neo-Hookean or Mooney-Rivlin material will be discussed in detail in section~\ref{sec:estable}. Next, we formulate the stochastic version of this problem.

\subsection{Stochastic setting}\label{sec:stochastic}

We now consider an ensemble of cubes, where each cube is made from a single homogeneous isotropic incompressible hyperelastic material, with the elastic parameters distributed according to a known probability function. For each individual cube, the finite elasticity theory applies. Nevertheless, in extending this theory to the ensemble, caution must be exercised when combining the nonlinearity of the hyperelastic models with the probability laws \cite{McCoy:1973}. Examples of different stochastic isotropic hyperelastic models derived from experiment are presented in \cite{Staber:2017:SG,Mihai:2018a:MWG}. Specifically, we assume that each cube is made of a stochastic Mooney-Rivlin material characterised by the constitutive law \cite{Staber:2015:SG}
\begin{equation}\label{eq:smr}
\mathcal{W}_{smr}(\lambda_{1},\lambda_{2},\lambda_{3})=\frac{\mu_{1}}{2}\left(\lambda_{1}^2+\lambda_{2}^2+\lambda_{3}^2-3\right)
+\frac{\mu_{2}}{2}\left(\frac{1}{\lambda_{1}^2}+\frac{1}{\lambda_{2}^2}+\frac{1}{\lambda_{3}^2}-3\right),
\end{equation}
where $\mu_{1}$ and $\mu_{2}$ are now given random parameters, to be described in terms of probability distributions. The model \eqref{eq:smr} simplifies as a stochastic neo-Hookean model if $\mu_{2}=0$. Consistent with the deterministic elastic definition \cite{Mihai:2017:MG}, the random shear modulus of the stochastic Mooney-Rivlin material \eqref{eq:smr} under small strain is equal to $\mu=\mu_{1}+\mu_{2}$ \cite{Staber:2015:SG,Mihai:2018a:MWG}. Then, assumption (A4) is guaranteed by the following constraints on the expected values \cite{Staber:2015:SG,Staber:2016:SG,Staber:2017:SG,Mihai:2018a:MWG}:
\begin{eqnarray}\label{eq:Emu1}\begin{cases}
E\left[\mu\right]=\underline{\mu}>0,&\\
E\left[\log\ \mu\right]=\nu,& \mbox{such that $|\nu|<+\infty$}.\label{eq:Emu2}\end{cases}
\end{eqnarray}
As shown in \cite{Soize:2000,Soize:2001}, under the above set of constraints (\ref{eq:Emu1}), the random shear modulus, $\mu$, follows a Gamma probability distribution \cite{Abramowitz:1964,Johnson:1994:JKB}, with hyperparameters $\rho_{1}>0$ and $\rho_{2}>0$ satisfying
\begin{equation}\label{eq:rho12}
\underline{\mu}=\rho_{1}\rho_{2},\qquad
\text{Var}[\mu]=\rho_{1}\rho_{2}^2,
\end{equation}
where  $\underline{\mu}$ is the mean value and $\text{Var}[\mu]$ is the variance of $\mu$. The corresponding probability density function takes the form
\begin{equation}\label{eq:gammamu}
g(\mu;\rho_{1},\rho_{2})=\frac{\mu^{\rho_{1}-1}e^{-\mu/\rho_{2}}}{\rho_{2}^{\rho_{1}}\Gamma(\rho_{1})},\qquad\mbox{for}\ \mu>0\ \mbox{and}\ \rho_{1}, \rho_{2}>0,
\end{equation}
where $\Gamma:\mathbb{R}^{*}_{+}\to\mathbb{R}$ is the complete Gamma function
\begin{equation}\label{eq:gamma}
\Gamma(z)=\int_{0}^{+\infty}t^{z-1}e^{-t}dt.
\end{equation}
Setting a fixed constant value $b>-\infty$, such that $\mu_{i}>b$, $i=1,2$ (e.g., $b=0$ if $\mu_{1}>0$ and $\mu_{2}>0$, although $b$ is not unique in general), we define the auxiliary random variable \cite{Mihai:2018a:MWG}
\begin{equation}\label{eq:R12:b}
R_{1}=\frac{\mu_{1}-b}{\mu-2b},
\end{equation}
such that $0<R_{1}<1$. Then, the random model parameters can be expressed equivalently as follows,
\begin{equation}\label{eq:mu12:b}
\mu_{1}=R_{1}(\mu-2b)+b,\qquad \mu_{2}=\mu-\mu_{1}=(1-R_{1})(\mu-2b)+b.
\end{equation}
It is reasonable to assume \cite{Staber:2015:SG,Staber:2016:SG,Staber:2017:SG,Mihai:2018a:MWG}
 \begin{eqnarray}\begin{cases}
E\left[\log\ R_{1}\right]=\nu_{1},& \mbox{such that $|\nu_{1}|<+\infty$},\label{eq:ER1}\\
E\left[\log(1-R_{1})\right]=\nu_{2},& \mbox{such that $|\nu_{2}|<+\infty$},\label{eq:ER2}\end{cases}
\end{eqnarray}
in which case, the random variable $R_{1}$ follows a standard Beta distribution \cite{Abramowitz:1964,Johnson:1994:JKB}, with hyperparameters $\xi_{1}>0$ and $\xi_{2}>0$ satisfying
\begin{equation}\label{eq:xi12}
\underline{R}_{1}=\frac{\xi_{1}}{\xi_{1}+\xi_{2}},\qquad
\text{Var}[R_{1}]=\frac{\xi_{1}\xi_{2}}{\left(\xi_{1}+\xi_{2}\right)^2\left(\xi_{1}+\xi_{2}+1\right)},
\end{equation}
where $\underline{R}_{1}$ is the mean value and $\text{Var}[R_{1}]$ is the variance of $R_{1}$. The associated probability density function is 
\begin{equation}\label{eq:betaR1}
\beta(r;\xi_{1},\xi_{2})=\frac{r^{\xi_{1}-1}(1-r)^{\xi_{2}-1}}{B(\xi_{1},\xi_{2})},\qquad \qquad\mbox{for}\ r\in(0,1)\ \mbox{and}\ \xi_{1}, \xi_{2}>0,
\end{equation}
where $B:\mathbb{R}^{*}_{+}\times\mathbb{R}^{*}_{+}\to\mathbb{R}$ is the Beta function
\begin{equation}\label{eq:beta}
B(x,y)=\int_{0}^{1}t^{x-1}(1-t)^{y-1}dt.
\end{equation}
Then, for the random coefficients given by \eqref{eq:mu12:b}, the corresponding mean values are
\begin{equation}\label{eq:mu12:mean}
\underline{\mu}_{1}=\underline{R}_{1}(\underline{\mu}-2b)+b,
\qquad \underline{\mu}_{2}=\underline{\mu}-\underline{\mu}_{1}=(1-\underline{R}_{1})(\underline{\mu}-2b)+b.,
\end{equation}
and the variances and covariance take the form, respectively,
\begin{eqnarray}
&&\text{Var}\left[\mu_{1}\right]
=(\underline{\mu}-2b)^2\text{Var}[R_{1}]+(\underline{R}_{1})^2\text{Var}[\mu]+\text{Var}[\mu]\text{Var}[R_{1}],\\
&&\text{Var}\left[\mu_{2}\right]
=(\underline{\mu}-2b)^2\text{Var}[R_{1}]+(1-\underline{R}_{1})^2\text{Var}[\mu]+\text{Var}[\mu]\text{Var}[R_{1}],\\
&&\text{Cov}[\mu_{1},\mu_{2}]=\frac{1}{2}\left(\text{Var}[\mu]-\text{Var}[\mu_{1}]-\text{Var}[\mu_{2}]\right).
\end{eqnarray}
As in the deterministic elastic case, each homogeneous isotropic hyperelastic cube is subject solely to equitriaxial dead-load tractions, uniformly distributed on all faces in the reference configuration, and every cube deforms by a homogeneous triaxial stretch. The question is: For a known dead load, given a distribution of homogeneous mechanical properties, what is the probability distribution of stable triaxial deformations? Before we can answer this question, it is instructive to review the solutions to the deterministic elastic problem, which forms the baseline for our stochastic elastic approach.

\section{Equilibria of an elastic cube}\label{sec:estable}

When an elastic body of homogeneous isotropic incompressible hyperelastic material deforms by $\textbf{x}(\textbf{X})$ under the sole action of the equitriaxial dead-load traction $\tau$, the \emph{total free energy} is equal to
\begin{equation}\label{eq:energy}
E(\textbf{x})=\int_{V}\left[W(\textbf{F})-\tau\text{tr}(\textbf{F})\right]dV,
\end{equation}
where $W(\textbf{F})$ is the strain-energy density function of the material, expressed in terms of the gradient tensor $\nabla\textbf{x}=\textbf{F}$. As we are interested in the minimisers of this energy for an incompressible material, we only consider deformation gradients that satisfy  the incompressibility constraint $\det\textbf{F}=1$. This incompressibility constraint can be easily enforced by considering the minimisers of the  following unconstrained form,
\begin{equation}\label{eq:uenergy}
E_{o}(\textbf{x})=\int_{V}\left[W(\textbf{F})-\tau\text{tr}(\textbf{F})-p\left(\det\textbf{F}-1\right)\right]dV,
\end{equation}
where $p$ is a Lagrange multiplier that is interpreted as the hydrostatic pressure. In the case of a homogeneous triaxial stretch $\textbf{x}(\textbf{X})$, with gradient tensor $\nabla\textbf{x}=\text{diag}(\lambda_{1},\lambda_{2},\lambda_{3})$, we  define the function
\begin{equation}\label{eq:psi}
\Psi(\lambda_{1},\lambda_{2},\lambda_{3};\tau)=\mathcal{W}(\lambda_{1},\lambda_{2},\lambda_{3})-\tau\left(\lambda_{1}+\lambda_{2}+\lambda_{3}\right),
\end{equation}
where $\mathcal{W}(\lambda_{1},\lambda_{2},\lambda_{3})$ is the strain-energy density function for the isotropic elastic material expressed in terms of the principal stretches $\{\lambda_{i}\}_{i=1,2,3}$. In this case, the total free energy, given by \eqref{eq:energy}, takes the equivalent form
\begin{equation}\label{eq:energy:psi}
E(\textbf{x})=\int_{V}\Psi(\lambda_{1},\lambda_{2},\lambda_{3};\tau)dV,
\end{equation}
and we are interested in the minimisers of this energy when the deformation gradients satisfy the incompressibility condition \eqref{eq:inc}. Alternatively, we introduce the incompressibility constraint by defining
\begin{equation}\label{eq:upsi}
\Psi_{o}(\lambda_{1},\lambda_{2},\lambda_{3};\tau)=\mathcal{W}(\lambda_{1},\lambda_{2},\lambda_{3})-\tau\left(\lambda_{1}+\lambda_{2}+\lambda_{3}\right)-p\left(\lambda_{1}\lambda_{2}\lambda_{3}-1\right),
\end{equation}
the corresponding total free energy given by \eqref{eq:uenergy} is equal to
\begin{equation}\label{eq:energy:upsi}
E_{o}(\textbf{x})=\int_{V}\Psi_{o}(\lambda_{1},\lambda_{2},\lambda_{3};\tau)dV.
\end{equation}

\begin{definition}\label{def:stable}
We say that the deformation $\textbf{x}=\textbf{x}(\textbf{X})$, with gradient $\nabla\textbf{x}$, such that $\det(\nabla\textbf{x})=1$, is ``stable'' if it is a local minimum of the total free energy, i.e., if the following inequality holds
	\begin{equation}\label{eq:stable}
	E(\textbf{x})< E(\textbf{y}),
	\end{equation}
for all $\textbf{y}=\textbf{y}(\textbf{X})$ that are continuous, piecewise differentiable deformation mappings, with gradients  $\nabla\textbf{y}$ satisfying $\det(\nabla\textbf{y})=1$. When relation (\ref{eq:stable}) holds with ``$\leq$'' instead of the strict inequality ``$<$'', the deformation is ``neutrally stable''. Otherwise, the deformation is ``unstable''.
\end{definition}

The following general result is central to the stability analysis in this paper (see theorem 2.2 of \cite{Ball:1983:BS} and the Appendix for a proof).

\begin{theorem}\label{th:stability}
When an homogeneous isotropic incompressible hyperelastic body, with the strain-energy density function $\mathcal{W}(\lambda_{1},\lambda_{2},\lambda_{3})$, deforms by a homogeneous triaxial stretch under the equitriaxial dead-load traction $\tau$, the following statements hold:
\begin{itemize}
\item[(i)] If $\tau<0$ (compressive loading), then only the trivial (undeformed, reference) configuration, with $\lambda_{1}=\lambda_{2}=\lambda_{3}=1$, is possible, and this state is unstable;
\item[(ii)] If $\tau=0$ (no loading), then the reference state is neutrally stable;
\item[(iii)] If $\tau>0$ (tensile loading), then the reference state is stable when it is a strict local minimum of $\Psi(\cdot,\tau)$, given by \eqref{eq:psi}, neutrally stable when it is a non-strict local minimum of $\Psi(\cdot,\tau)$, and unstable otherwise;
\item[(iv)] If $\tau>0$ (tensile loading), then any non-trivial local minimum of $\Psi(\cdot,\tau)$ is neutrally stable.
\end{itemize}
\end{theorem}

In particular, for a cube of Mooney-Rivlin material defined by \eqref{eq:mr}, the reference state is stable if $0<\tau<\tau_{0}=2\mu=2(\mu_{1}+\mu_{2})$, and a bifurcation occurs when $\tau=\tau_{0}=2\mu$. Then, as the dead load increases, the reference state becomes unstable, and neutrally stable non-trivial homogeneous triaxial deformations are possible. Explicitly, Mooney-Rivlin models with $\mu_{1}>0$ and $\mu_{2}\geq 0$, suitable for describing rubberlike materials, were treated in \cite{Ball:1983:BS}. However, the general stability analysis developed there is applicable also to the case when $\mu_{1}>0$ and $\mu_{2}<0$, given that the shear modulus satisfies $\mu=\mu_{1}+\mu_{2}>0$. For rubber, some negative values of $\mu_{2}$ were obtained in \cite{Kearsley:1980:KZ} from experimental data reported in \cite{Penn:1976:PK}. More generally, we recall that, for a cube of Mooney-Rivlin material subject to simple shear, if $\mu_{2}>0$, then the positive Poynting effect occurs, and if $-\mu_{1}<\mu_{2}<0$, then the negative Poynting effect is obtained. When $\mu_{2}=0$, the model takes the neo-Hookean form and there is no Poynting effect. The (positive or negative) \emph{Poynting effect} is a large strain effect observed when an elastic cube is sheared between two plates and stress is developed in the direction normal to the sheared faces, or when a cylinder is subjected to torsion and the axial length changes \cite{Poynting:1909,Truesdell:1952,Rivlin:1953,Moon:1974:MT,Janmey:2006:JMRLGM,Destrade:2011:DMS,Mihai:2011:MG,Mihai:2013:MG,Mihai:2017:MG}. Here, we  treat separately the cases of one-term models, with $\mu_{2}=0$, and  two-term models, with $\mu_{2}\neq0$, respectively.

\subsection{The neo-Hookean cube} When $\mu_{2}=\mu-\mu_{1}=0$ and $\mu=\mu_{1}>0$, the Mooney-Rivlin model given by \eqref{eq:mr} reduces to a (one-term) neo-Hookean form \cite{Treloar:1944}. This is a well-known case and we briefly summarise the results for further comparison. For $\tau>\tau_{*}=3\mu/2^{2/3}$, there exist, in addition to the trivial state, two more equilibria. At $\tau=2\mu$, the trivial solution becomes unstable and there is a bifurcation into a plate-like deformation (one stretch less than $1$ and two equal stretches greater than $1$) and a rod-like deformation (one stretch greater than $1$ and two equal stretches less than $1$) \cite{Rivlin:1974,Ball:1983:BS} (figure~\ref{rivlincube}). By rotation, each of these solutions define three possible deformations for a total of six non-trivial deformations for $\tau>\tau_{*}$ (with $\tau\not<2\mu$).

\begin{figure}[htbp]
	\begin{center}
		\includegraphics[width=0.8\textwidth]{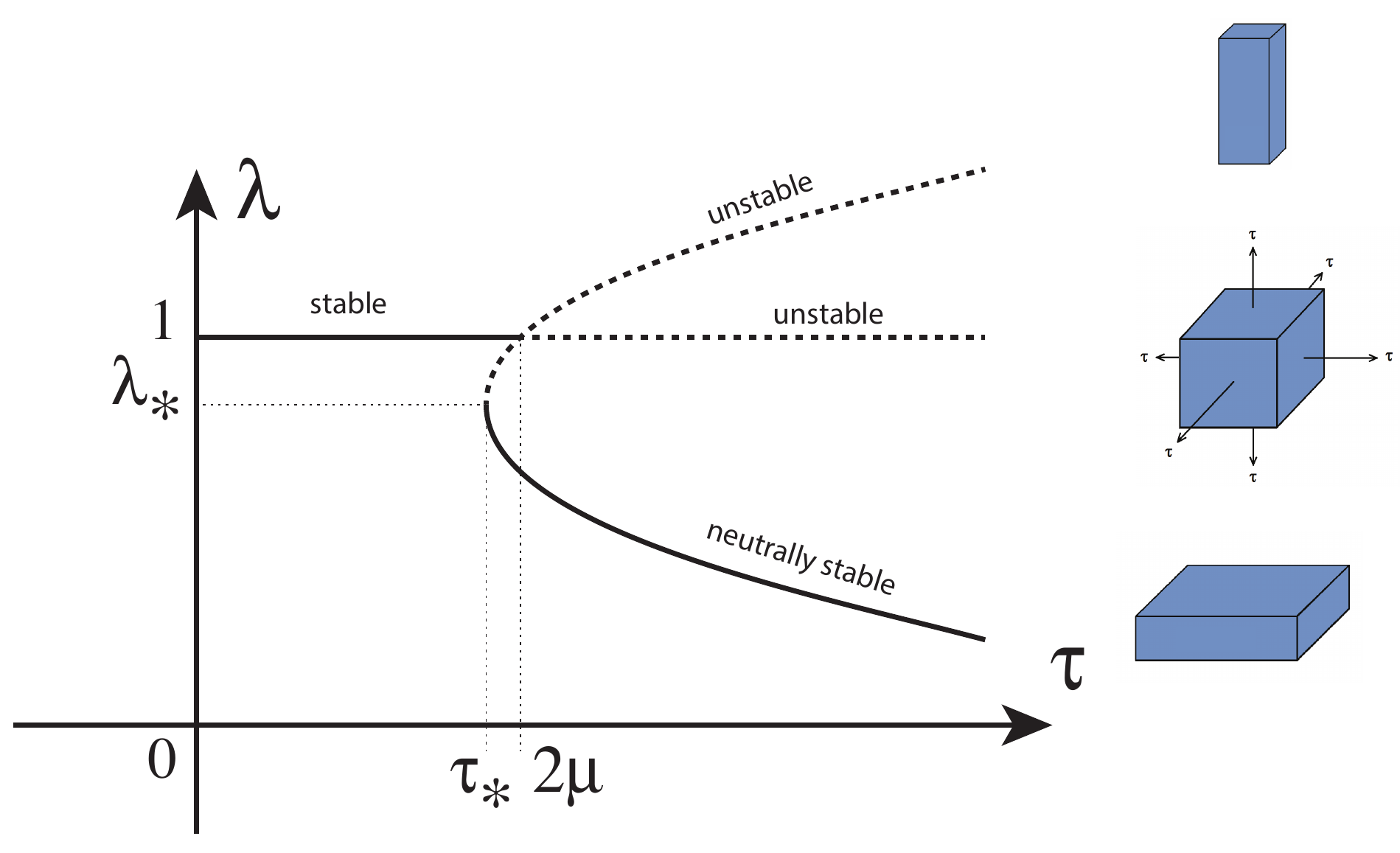}
		\caption{Bifurcation diagrams showing possible homogeneous equilibria of neo-Hookean  cube under normal dead-load tractions, $\tau>0$, uniformly distributed on all faces in the reference configuration.}\label{nh-eldiag}
	\end{center}
\end{figure}

For the non-trivial homogeneous triaxial stretches, $0<\lambda\neq 1$ is a solution of the equation
	\begin{equation}\label{eq:cubic}
	\lambda^{3/2}-\frac{\tau}{\mu}\lambda^{1/2}+1=0.
	\end{equation}
Without loss of generality, we label the axis such that the rod-like solution is obtained by taking $\lambda_{1}=\lambda>1$, and $\lambda_{2}=\lambda_{3}=1/\sqrt{\lambda}<1$. The plate-like solution is obtained by taking $\lambda_{1}=\lambda<1,$ $\lambda_{2}=\lambda_{3}=1/\sqrt{\lambda}>1$. The deformations are neutrally stable if $\lambda<\tau/(3\mu)$, and unstable if $\lambda>\lambda_{*}=\tau/(3\mu)$. The corresponding dead loads take the form
\begin{equation}\label{eq:tau:nh}
\tau=\mu\left(\lambda+\frac{1}{\sqrt{\lambda}}\right)>0,
\end{equation}
hence, non-positive loads are not possible. When $\tau=\tau_{*}=3\mu/2^{2/3}$, only one non-trivial solution for the cubic equation \eqref{eq:cubic} exists, namely $\lambda=\lambda_{*}=\tau_{*}/(3\mu)=1/2^{2/3}$, and the corresponding plate-like deformation is neutrally stable. By rotation, there are three such non-trivial deformations. The behavior of the neo-Hookean cube under equitriaxial tensile tractions is illustrated in figure~\ref{nh-eldiag}.

\subsection{The Mooney-Rivlin cube} When $0<\mu=\mu_{1}+\mu_{2}\neq\mu_{1}$, we separate the cases where $\mu_{2}>0$ and $\mu_{2}<0$, respectively.

\begin{figure}[htbp]
	\begin{center}
		\includegraphics[width=0.7\textwidth]{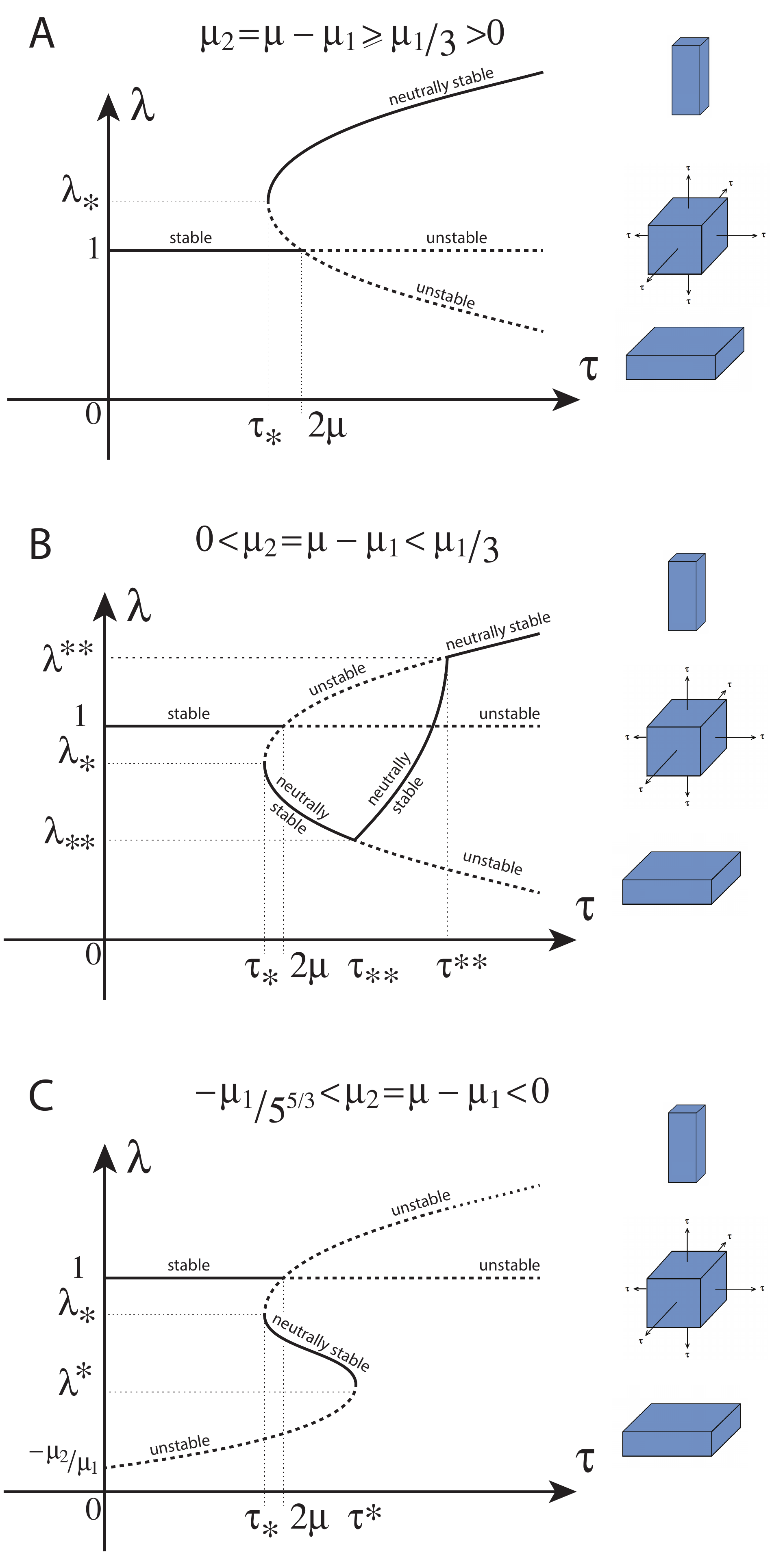}
		\caption{Bifurcation diagrams showing possible homogeneous equilibria of Mooney-Rivlin cubes under normal dead-load tractions.}\label{mr-eldiag}
	\end{center}
\end{figure}

\subsubsection{Positive $\mu_2$} 
If $0<\mu_{2}=\mu-\mu_{1}<\mu$, then, under sufficiently large tensile loading, the reference state is unstable, and there is a bifurcation into rod-like and plate-like deformations \cite{Ball:1983:BS}. By rotation, there are six non-trivial deformations, with the corresponding dead load taking the form
\begin{equation}\label{eq:tau:mr}
\tau=\left(\mu_{1}+\frac{\mu_{2}}{\lambda}\right)\left(\lambda+\frac{1}{\sqrt{\lambda}}\right)>0,
\end{equation}
hence, non-positive loads are not possible. Specifically:
	
\paragraph{E1.1.} When $\mu_{2}=\mu-\mu_{1}\geq\mu_{1}/3>0$, neutrally stable rod-like configurations are possible, such that $\lambda>\lambda_{*}$, for some $\lambda_{*}\geq1$, while all other configurations are unstable (figure~\ref{mr-eldiag}A). If $\mu_{2}=\mu_{1}/3>0$, then $\lambda_{*}=1$ and rod-like configurations are neutrally stable while plate-like ones are unstable.

\paragraph{E1.2.} When $0<\mu_{2}=\mu-\mu_{1}<\mu_{1}/3$, under sufficiently small dead loads, plate-like states are neutrally stable, while rod-like ones are unstable. When the dead load is increased, a secondary bifurcation into a neutrally stable deformation with three unequal stretches, $\lambda_{1}\neq\lambda_{2}\neq\lambda_{3}\neq\lambda _{1}$, that links the rod-like branch to the plate-like one is possible, after which rod-like deformations are neutrally stable (figure~\ref{mr-eldiag}B). Indeed, in the case of three unequal stretches, by \eqref{eq:pki:mr} and \eqref{eq:tractions:mr},
\begin{equation}
\tau=\left(\mu_{1}+\mu_{2}\lambda_{1}^2\right)\left(\lambda_{2}+\lambda_{3}\right)=\left(\mu_{1}+\mu_{2}\lambda_{3}^2\right)\left(\lambda_{1}+\lambda_{2}\right).
\end{equation}
Next, eliminating $\tau$ from the above identities implies
\begin{equation}\label{eq:muslambdas}
\mu_{1}=\mu_{2}\left(\lambda_{1}\lambda_{2}+\lambda_{2}\lambda_{3}+\lambda_{3}\lambda_{1}\right).
\end{equation}
Then, by \eqref{eq:inc} and \eqref{eq:muslambdas}, the dead load takes the form
\begin{equation}\label{eq:tau123}
\tau=\frac{\mu_{2}}{\lambda_{3}}\left(\frac{\mu_{1}}{\mu_{2}}+\lambda_{3}^2\right)\left(\frac{\mu_{1}}{\mu_{2}}-\frac{1}{\lambda_{3}}\right),
\end{equation}
and has a maximum, $\tau^{**}>0$, where it intersects the rod-like branch, with $\lambda=\lambda^{**}>1$ and a minimum, $\tau_{**}>0$, where it intersects the plate-like branch, with $\lambda=\lambda_{**}<1$. The points of intersection can be found by setting $\lambda_{3}=1/\sqrt{\lambda}$ in \eqref{eq:tau123}.

\subsubsection{Negative $\mu_2$} 
 If $-\mu_{1}<\mu_{2}=\mu-\mu_{1}<0$, then we recall that a primary kinematic assumptions is that the Baker-Ericksen (BE) inequalities \eqref{eq:BE} hold \cite[p.~158]{TruesdellNoll:2004}. Combining \eqref{eq:sigmai:mr} and \eqref{eq:BE} gives
\begin{equation}\label{eq:BE:mr}
\mu_{1}+\mu_{2}\lambda_{k}^2>0\quad \mbox{if}\quad \lambda_{i}\neq\lambda_{j},\ i\neq k\neq j,\quad i,j,k=1,2,3.
\end{equation}
Clearly, the above inequalities are satisfied if $\mu_{1}>0$ and $\mu_{2}\geq 0$. When $0>\mu_{2}>-\mu_{1}$, these inequalities are equivalent to
\begin{equation}\label{eq:BE:lambda}
0<\lambda_{k}^2<-\frac{\mu_{1}}{\mu_{2}}\quad \mbox{if}\quad \lambda_{i}\neq\lambda_{j},\quad i\neq k\neq j,\quad i,j,k=1,2,3.
\end{equation}
Assuming the deformation with three unequal stretches, $\lambda_{1}\neq\lambda_{2}\neq\lambda_{3}\neq\lambda _{1}$, by \eqref{eq:tau123} and \eqref{eq:BE:mr}, $\tau>0$. However, as $\mu_{1}>0$ and $\lambda_{i}>0$, $i=1,2,3$, \eqref{eq:muslambdas} implies that three unequal stretches are impossible when $\mu_{2}<0$. Therefore, only non-trivial deformations with two equal stretches (rod-like or plate-like) require further consideration. In this case, setting $\lambda_{k}=1/\sqrt{\lambda}$ in \eqref{eq:BE:lambda} implies $\lambda>-\mu_{2}/\mu_{1}$, and the dead load, $\tau$, takes the form given by \eqref{eq:tau:mr}. As $0<-\mu_{2}/\mu_{1}<1$, both $\lambda>1$ (rod-like states) and $\lambda<1$ (plate-like states) are possible. Direct calculations further reveal that:

\paragraph{E2.1.} When $-\mu_{1}/5^{5/3}<\mu_{2}<0$, there is a maximum dead load, $\tau^{*}>0$, attained for some $\lambda^{*}$, satisfying $-\mu_{2}/\mu_{1}<\lambda^{*}<-5\mu_{2}/\mu_{1}<1/5^{2/3}$, and a minimum dead load, $\tau_{*}>0$, attained for some, $\lambda_{*}$, such that $-5\mu_{2}/\mu_{1}<\lambda_{*}<1$. In this case, neutrally stable plate-like solutions are possible under dead loads satisfying $\tau_{*}<\tau<\tau^{*}$ (figure~\ref{mr-eldiag}C). For example, taking $\mu_{1}=2.484$ and $\mu_{2}=-0.148$, as reported in \cite{Kearsley:1980:KZ}, we obtain $\mu_{2}/\mu_{1}=-0.0596>-1/5^{5/3}\approx -0.0684$, thus such possibilities are not unrealistic. When $\tau^{*}>\tau_{0}=2\mu$, neutrally stable plate-like deformations are obtained after the reference state becomes unstable. This situation occurs if
\begin{equation}\label{eq:infimum}
\inf_{0<\lambda<1}\frac{\lambda^{5/2}-2\lambda^{3/2}+\lambda}{\lambda^{3/2}-1}\approx -0.045<\frac{\mu_{2}}{\mu_{1}}<0,
\end{equation}
where '$\inf$' denotes infimum.

\paragraph{E2.2.} When $\mu_{2}/\mu_{1}\leq -1/5^{5/3}<0$, the non-trivial homogeneous deformations are unstable.

\section{Probability distribution of equilibria of a stochastic cube}\label{sec:sestable}

We now turn our attention to the stochastic problem introduced in section~\ref{sec:stochastic}. From the deterministic elastic problem, we note that, in the absence of loading, $\tau=0$, the reference state is neutrally stable, i.e., the identity deformation gradient is a minimum for the elastic energy of each Mooney-Rivlin cube in the stochastic set. Further, under compressive dead-load traction, $\tau<0$, the reference state is unstable, and there are no other homogeneous triaxial states for these cubes. Thus, when $\tau\leq 0$, there is no uncertainty to be resolved about the reference state or its expected stability. Thus, we only need to consider here the case when the surface dead load is tensile, $\tau>0$. In this case, we derive the probability distribution of the homogeneous triaxial deformations for which the stochastic cube is in stable equilibrium. We do this by specifying the magnitude of the dead load and the probability distributions of the hyperelastic material parameters.

\subsection{The stochastic neo-Hookean cube}
\begin{figure}[htbp]
	\begin{center}
		\includegraphics[width=0.99\textwidth]{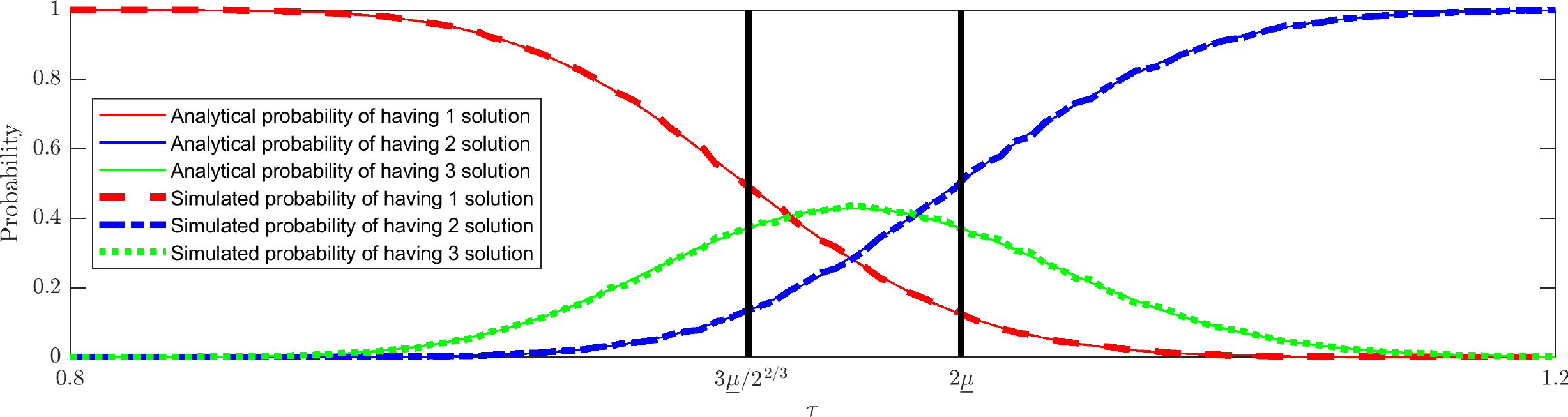}
		\caption{Calculated analytical (continuous lines) and numerical (dashed lines) probability distributions of different numbers of possible equilibria for a stochastic neo-Hookean cube with random shear modulus, $\mu>0$, taken from the Gamma distribution with $\rho_1=400$, $\rho_2=0.0013$. The two black lines delineate the expected regions of equilibria based only on the mean value of the shear modulus, $\underline{\mu}=\rho_1\rho_2$.}\label{nh-intpdfs}
	\end{center}
\end{figure}

\begin{figure}[htbp]
	\begin{center}
		A\includegraphics[width=0.45\textwidth]{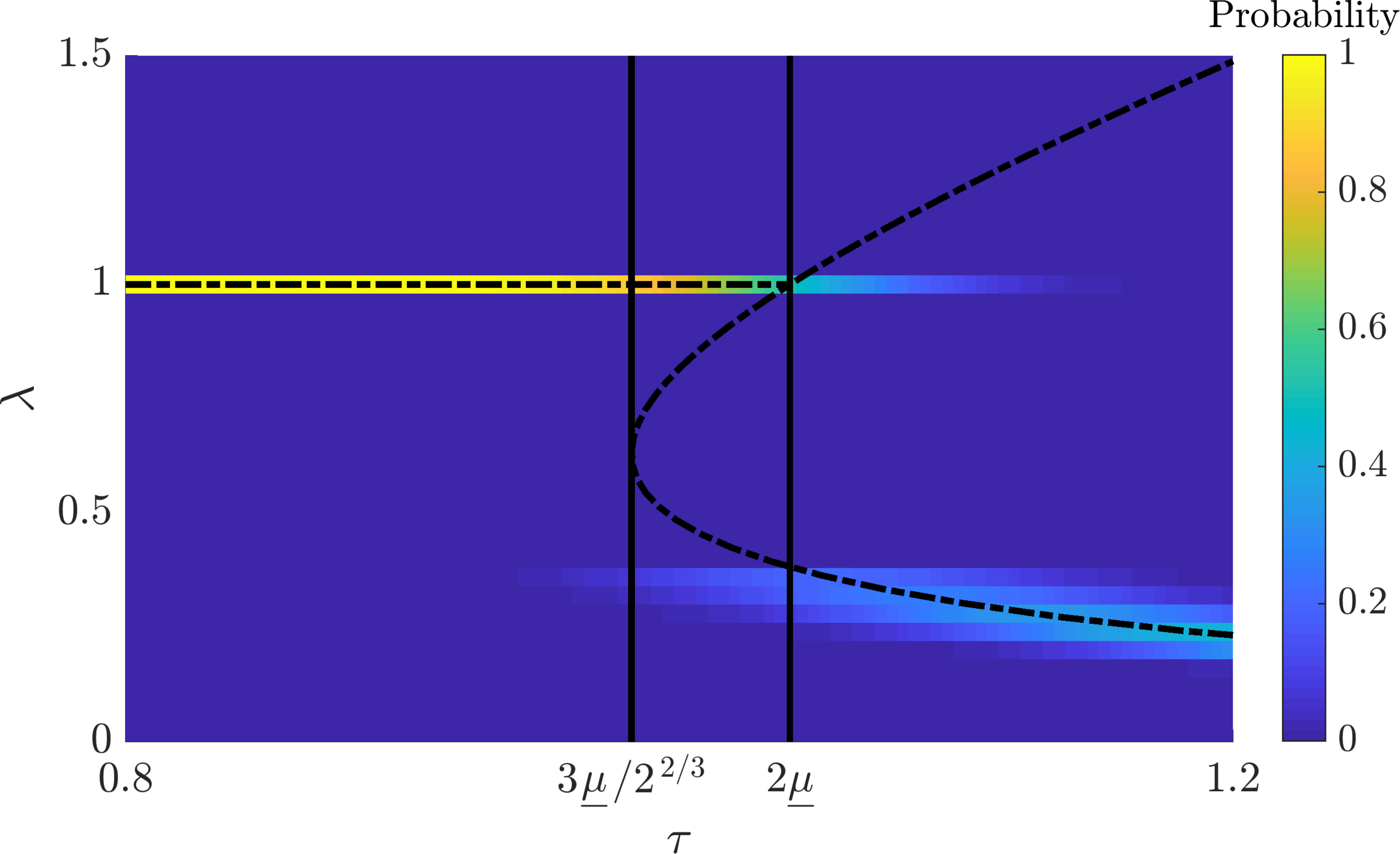}\qquad
		B\includegraphics[width=0.42\textwidth]{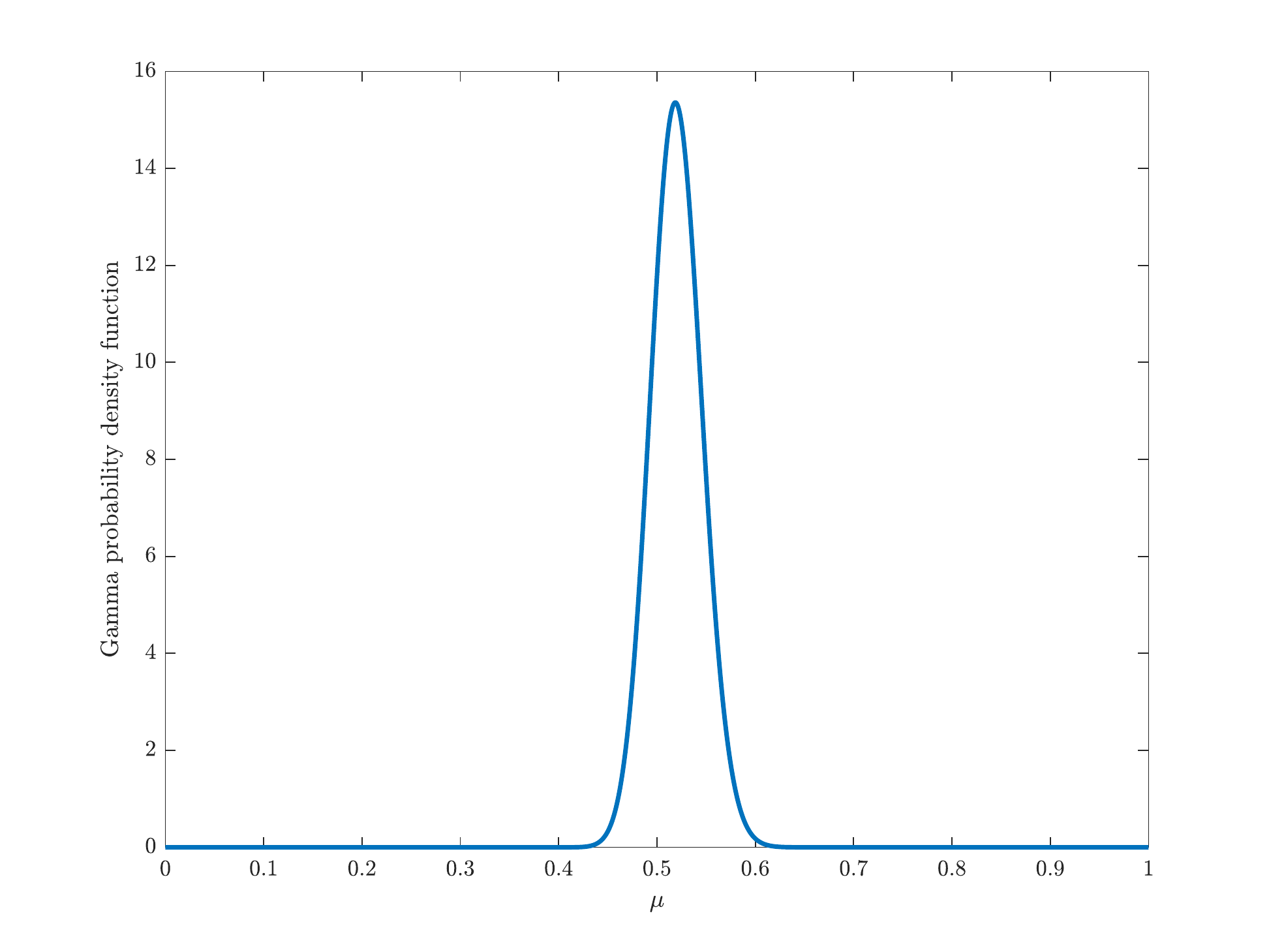}
		\caption{(A) Stochastic bifurcation diagram for the probability distribution of stretches as function of dead load, $\tau$, for a stochastic neo-Hookean cube with $\rho_1=400$, $\rho_2=0.0013$. The probability distribution follows the stable branches of the bifurcation diagram (see also figure~\ref{nh-eldiag}). The two black solid lines delineate the expected regions of equilibria based solely on the mean value of the shear modulus, $\underline{\mu}=\rho_1\rho_2$. (B) Assumed Gamma distribution with $\rho_1=400$, $\rho_2=0.0013$ for the random shear modulus, $\mu>0$.}\label{nh-stdiag}
	\end{center}
\end{figure}

For the stochastic -neo-Hookean cube, first, we look at the probability distributions for the number of equilibria as a function of $\tau$, given that $0<\mu$ follows a Gamma probability density function, $g\left(\mu,\rho_{1},\rho_{2}\right)$, given by \eqref{eq:gammamu}. As in the deterministic elastic case, under a dead-load traction $\tau>0$, there are three possible outcomes (figure~\ref{nh-eldiag}),
\begin{enumerate}
\item $\tau<3\mu/2^{2/3}$, or equivalently, $\mu>2^{2/3}\tau/3$, in which case the trivial, reference state is the unique stable state;\label{Trivial}
\item $0<2\mu<\tau$, or equivalently, $0<\mu<\tau/2$, in which case there are two non-trivial solutions, with $\lambda\neq 1$ satisfying equation \eqref{eq:tau:nh};
\item $3\mu/2^{2/3}<\tau<2\mu$, or equivalently, $\tau/2<\mu<2^{2/3}\tau$, in which case there are three possible equilibria, namely both the previous cases co-exist.\label{Non-trivial}
\end{enumerate}
Depending on the applied dead load, the probability of having one, two, or three possible equilibria are denoted by $P_1(\tau)$, $P_2(\tau)$ and $P_3(\tau)$, respectively, where
\begin{align}
P_1(\tau)&=1-\int_{0}^{2^{2/3}\tau/3}g(u;\rho_{1},\rho_{2})du,\label{P_1}\\
P_2(\tau)&=\int_0^{\tau/2}g(u;\rho_{1},\rho_{2})du,\label{P_2}\\
P_3(\tau)&=1-P_1(\tau)-P_2(\tau)=\int_{\tau/2}^{2^{2/3}\tau/3}g(u;\rho_{1},\rho_{2})du.\label{P_3}
\end{align}
To illustrate this numerically, we assume that shear modulus, $\mu>0$, follows a Gamma probability density function, with hyperparameters $\rho_1=400$, $\rho_2=0.0013$ (see figure~\ref{nh-stdiag}B). The resulting probability distributions given by equations \eqref{P_1}-\eqref{P_3} are illustrated in figure~\ref{nh-intpdfs} and compared with the distribution generated from stochastic simulations. Namely, the dead-load interval of $(0.8,1.2)$ was divided into $100$ steps, then for each value of $\tau$, $10^4$ random values of $\mu=\mu_1$ were numerically generated from a specified Gamma distribution and then compared with the inequalities defining the three intervals for values of $\tau$. Finally, from tallying the number of $\mu$ values that fall into each category, we are able to calculate the approximate corresponding probabilities, $P_1$, $P_2$ and $P_3$, and compare them to their respective analytical probabilities obtained exactly from \eqref{P_1}-\eqref{P_3}.

Figure~\ref{nh-intpdfs} can be understood as follows: Suppose that we consider a cube with the mechanical properties defined by the average of the Gamma distribution, $\underline{\mu}$, and we specify that the dead load is $\tau=2\underline{\mu}=2\rho_1\rho_2=1.04$. In the deterministic case, this is a bifurcation point where the cube transitions from having three equilibria (one trivial, two non-trivial) to having only two non-trivial equilibria. However, in the stochastic case, there is approximately: $10\%$ chance of a randomly chosen cube presenting only the single trivial state (red solid and dashed lines); $50\%$ chance of a randomly chosen cube presenting two non-trivial sates (blue solid and dashed lines); and $40\%$ chance of randomly choosing a cube that will present all three equilibria (green solid and dashed lines). In order to ensure that only the trivial equilibrium exists ($P_1\approx 1$), or to ensure that only  non-trivial equilibria exist ($P_2\approx 1$) we must look at values of the dead load beyond the expected bifurcation points. Namely, $P_1\approx 1$ when  $\tau\approx 0.8< 3\underline{\mu}/2^{2/3}$ and $P_2\approx 1$ when  $\tau\approx 1.2> 2\underline{\mu}$. Hence, accounting for the variability in the cube population has allowed us to predict that multiple equilibria will exist in a much larger region of the parameter space.

However, as in the deterministic elastic case, although multiple equilibria may exist, not all of the possible states may be stable. For the deterministic elastic problem, the stable branches are noted on the bifurcation diagram of figure~\ref{nh-eldiag}. Using this information, we are able to derive simulated probability distributions of what stretches are actually seen in the stochastic case. Specifically, during the $10^4$ stochastic trials, instead of coarsely categorising how many equilibria are possible, we use the generated value of $\mu$ and the given value of $\tau$ to calculate the observed stable stretch values. Critically, from figure~\ref{nh-eldiag}, we note that there are two possible cases. First, there is a single stable solution branch for the given value of $\tau$, either the trivial, reference configuration or the non-trivial configuration with smallest $\lambda$. In this case, the calculation of the stable solution is unique. Alternatively, in the case where there are multiple stable branches, one of the branches is always the trivial reference state. In this case, we  assume that the reference state is chosen. This assumption is based on the fact that the dead load is added to the reference state and, thus, unless perturbed in some way, it will remain in this state. 

The stable equilibria are illustrated in the \emph{stochastic bifurcation diagram} shown in figure~\ref{nh-stdiag}A, where $\rho_1=400$, $\rho_2=0.0013$. Hence $\underline{\mu}=\rho_1\rho_2=0.52$, and we can see that the probability distribution follows the stable branches of the bifurcation diagram (compare with figure~\ref{nh-eldiag}). Moreover, figure~\ref{nh-stdiag}B shows that the probability of the cube presenting the trivial reference state is approximately $50\%$, which matches the probability derived above. Namely, the probability that the trivial state, with $\lambda=1$, is observed is the probability that there is only one equilibrium, or that there are three equilibria, i.e., $P_1+P_3\approx 50\%$. The complementary $50\%$ appears as a distribution around $\lambda\approx 0.38$, which is the stable non-trivial stretch for the given $\tau$ and Gamma-distributed $\mu$.

In summary, for a stochastic neo-Hookean cube, under uniform tensile dead loads, given a shear modulus taken from a known Gamma probability distribution, we obtain the probabilities of stable equilibrium states. For the deterministic elastic problem, which is based on mean parameter value, there is a single-valued critical load that strictly separates the cases where either the trivial, reference configuration or the stable plate-like configuration occurs. By contrast, for the stochastic problem, there is a probabilistic load interval, containing the deterministic critical value, where there is a quantifiable chance for both the reference and plate-like states to be found.

\subsection{The stochastic Mooney-Rivlin cube}
Next, we  assume that $\mu_2=\mu-\mu_{1}\neq 0$, and consider the two possible cases, $\mu_2>0$ and $\mu_2<0$, respectively. When $\mu_2$ is negative, non-trivial equilibria only exist if $-\mu_1/5^{5/3}<\mu_2<0$. In all of these cases, as seen in section~\ref{sec:estable}, the probability of the trivial reference state being a stable equilibrium is exactly $1$ when $0<\tau<2\mu$, or equivalently, when $\mu>\tau/2>0$. Thus, if we specify a dead-load traction, $\tau$, then
\begin{equation}
P_0(\tau)=1-\int^{\tau/2}_{0}g(u;\rho_1,\rho_2)du
\end{equation}
is the probability that a cube chosen at random will present the trivial stable state. However, if $\tau>2\mu$, then the stable state will have to be calculated based on randomly generated values of $\mu$ and $\mu_1$, and the given value of $\tau$.

\subsubsection{Positive $\mu_2$} \label{positive_mu2}
To ensure that $\mu_2$ is positive, we set $b=0$ in equation \eqref{eq:R12:b} and define the Beta-distributed random variable
\begin{equation}\label{R1_mu2_positive}
R_{1}=\frac{\mu_{1}}{\mu}.
\end{equation}
To simulate the probability distribution of stable stretches, we fix a value of $\tau$ and simulate $10^4$ values of the Gamma distribution, $g(\mu;\rho_1,\rho_2)$, given by \eqref{eq:gammamu}, for $\mu$, and $10^4$ values of the Beta distribution, $\beta(r,\xi_1,\xi_2)$, given by \eqref{eq:betaR1}, for $R_{1}$. Then, we  calculate $\mu_1$ from \eqref{R1_mu2_positive}, and $\mu_2=\mu-\mu_1$ and, hence, specify which stability category the simulated cube's properties will fall into. For each random $\mu_1$ and $\mu_2$, we then find $\lambda$ from the corresponding algebraic equations of elastic equilibrium.

\begin{figure}[htbp]
	\begin{center}
		A\includegraphics[width=0.45\textwidth]{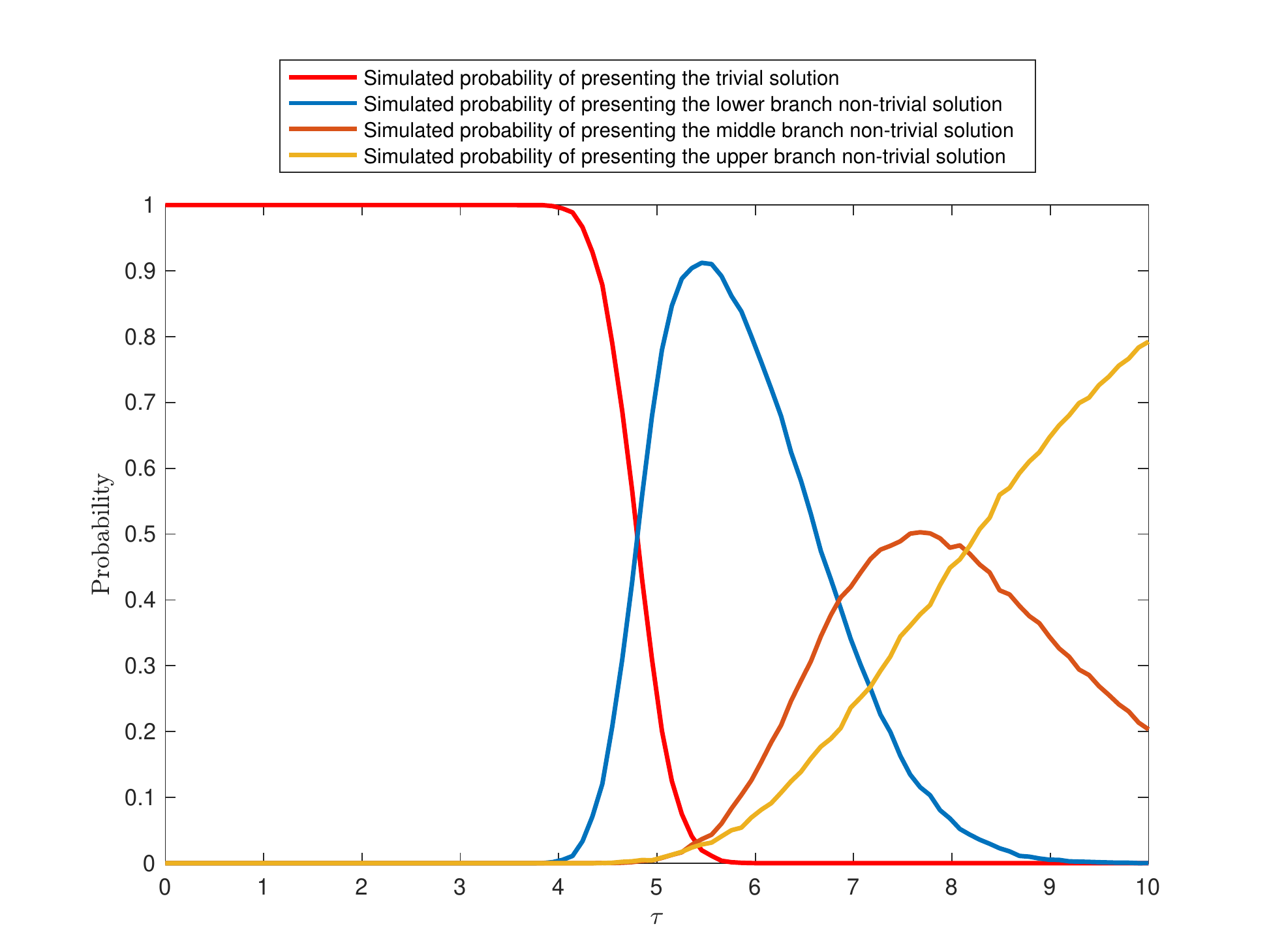}
		B\includegraphics[width=0.45\textwidth]{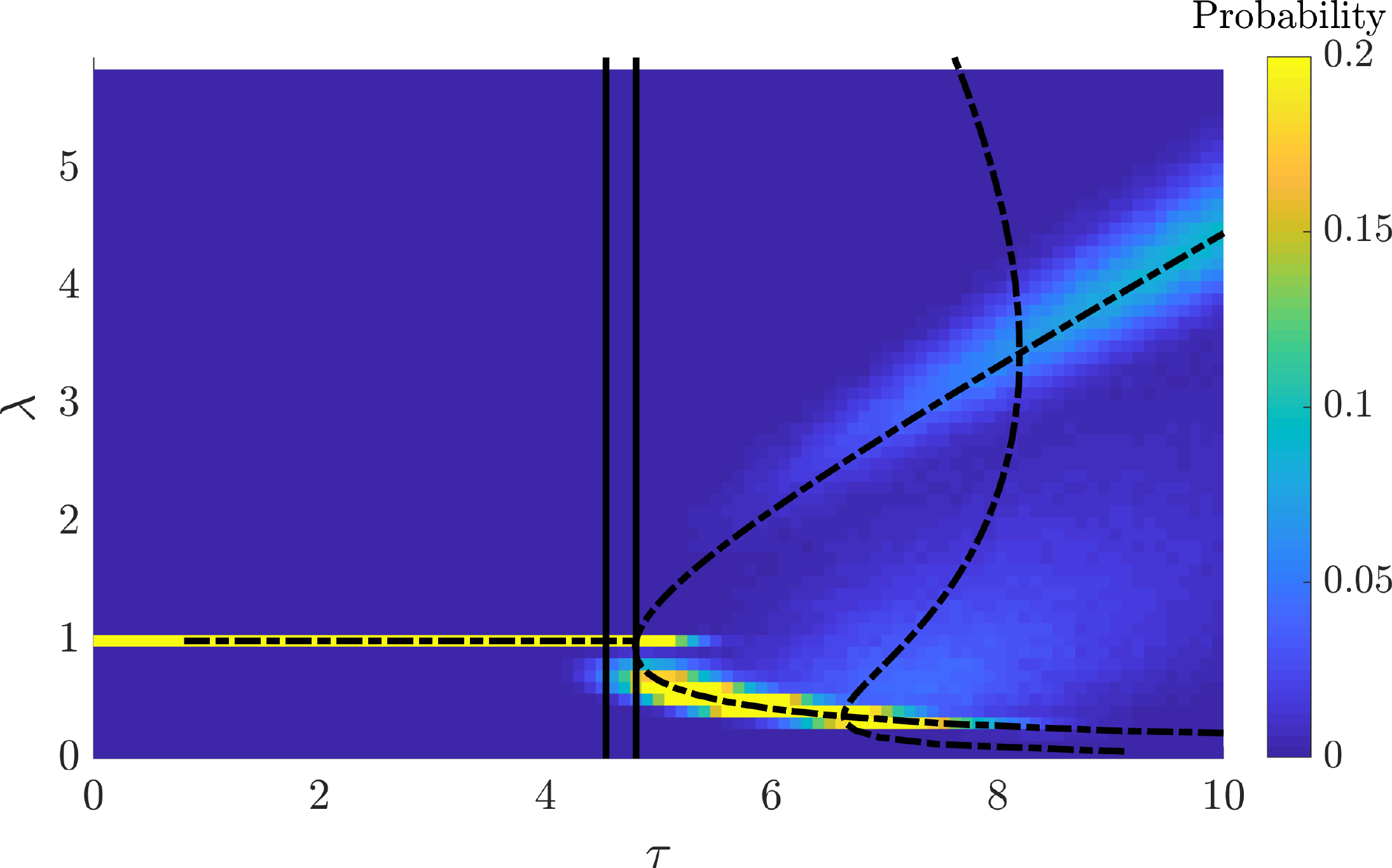}
		\caption{(A) Calculated probability distribution of possible equilibria for a stochastic Mooney-Rivlin cube with $0<\mu_{2}<\mu_{1}/3$ and $\rho_1=240$, $\rho_2=0.01$, $\xi_1=400$, $\xi_2=100$; (B) Stochastic bifurcation diagram with the probability distribution of stretches as function of dead load, $\tau$, following the stable branches of the diagram (see also figure~\ref{mr-eldiag}B).}\label{mr-mu2-positive-stdiag}
	\end{center}
	\begin{center}
		A\includegraphics[width=0.45\textwidth]{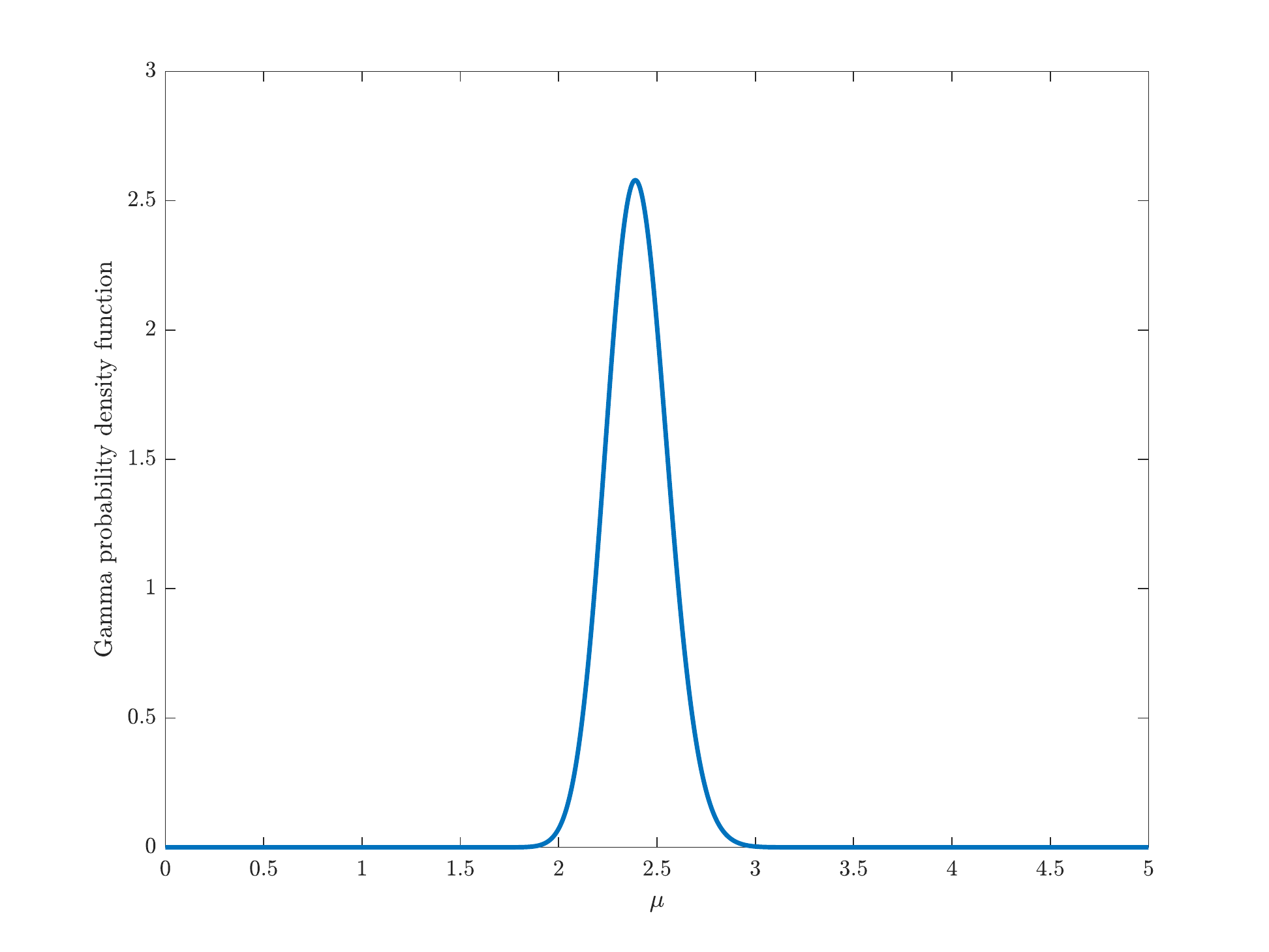}
		B\includegraphics[width=0.45\textwidth]{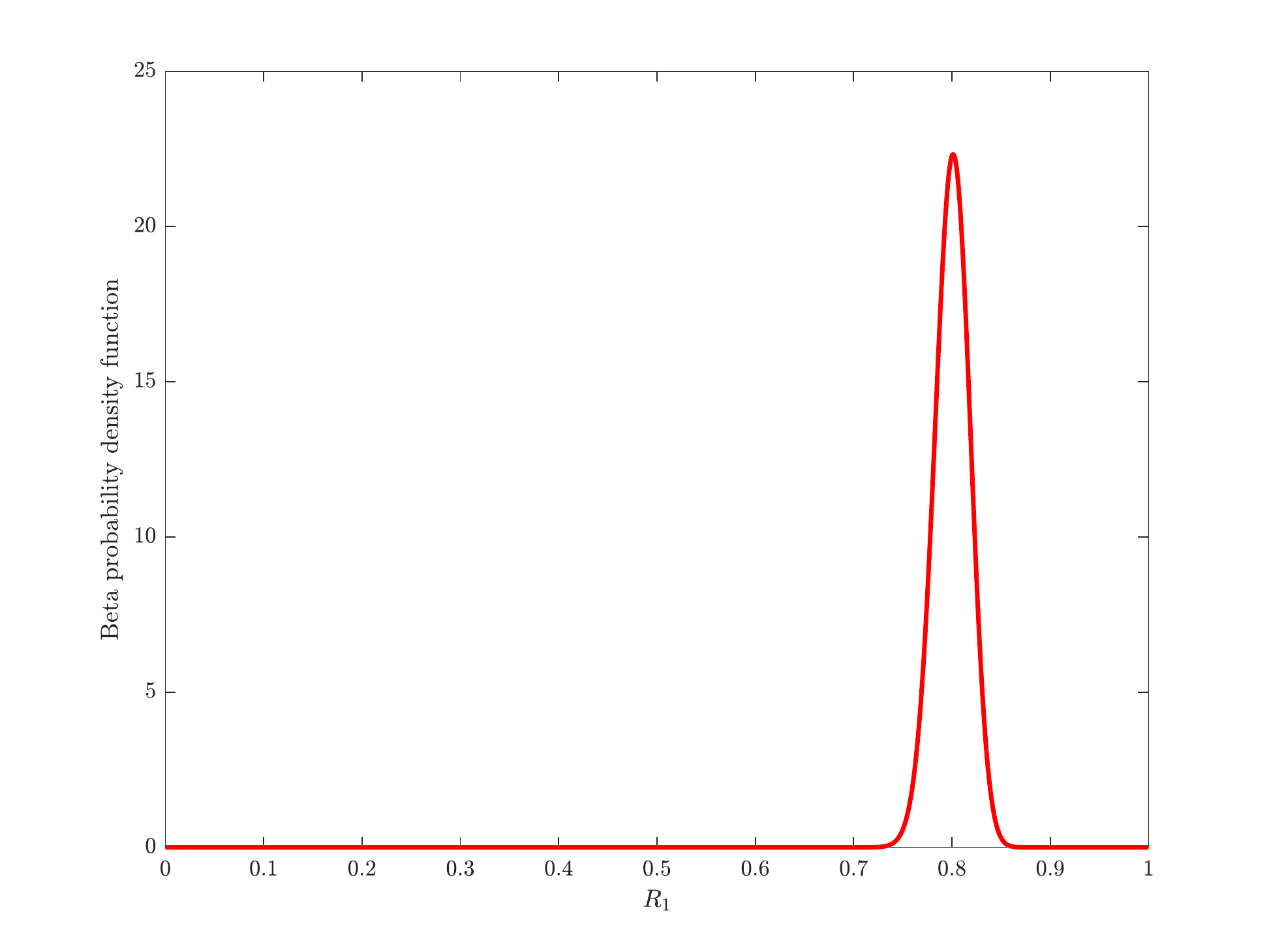}
		\caption{Assumed (A) Gamma distribution with $\rho_1=240$, $\rho_2=0.01$ for random shear modulus, $\mu=\mu_{1}+\mu_{2}>0$, and (B) Beta distribution with $\xi_1=400$, $\xi_2=100$ for random variable $R_{1}$ given by \eqref{R1_mu2_positive}, in the case of a stochastic Mooney-Rivlin cube with $0<\mu_{2}<\mu_{1}/3$.}\label{mr-mu2-positive-gbpdfs}
	\end{center}
\end{figure}

Note that, unless we further restrict the Beta distribution, there is a second bifurcation depending on the sign of $\mu_2-\mu_1/3$ (see figures~\ref{mr-eldiag}A and B). In the most general case, when we have no information about the mechanical properties of the cube, we cannot restrict the sign of $\mu_2-\mu_1/3$. Thus, due to the stochastic nature of $\mu_1$ and $\mu_{2}$, we cannot guarantee, a priori, the sign of $\mu_2-\mu_1/3$. However, we can calculate the probability of a cube presenting a given sign. Namely, $\mu_2-\mu_1/3$ is negative if and only if $R_1>3/4$, and  positive otherwise. Hence,
\begin{align}
P(\mu_2<\mu_1/3)&=\int^{1}_{3/4}\beta(r;\xi_1,\xi_2)dr,\\
P(\mu_2>\mu_1/3)&=\int^{3/4}_0\beta(r;\xi_1,\xi_2)dr.
\end{align}
By deriving appropriate limits in terms of intervals of $\mu_1$ and $\mu_{2}$, we are able to calculate the probability of a cube presenting a trivial or a specific non-trivial stable state within a given dead-load interval. However, it is much more useful to consider the stretches, $\lambda$, for a given value of $\tau$, in which case we invoke stochastic simulations.

Using the strategy of finding stable stretches, as discussed above, we are able to produce a \emph{stochastic bifurcation diagram}. For example, if $\rho_1=240$, $\rho_2=0.01$ and $\xi_1=400$, $\xi_2=100$, giving $\mu_2-\mu_1/3<0$ (see figure~\ref{mr-mu2-positive-gbpdfs}), then the expected values are $\underline{\mu}=\rho_1\rho_2=2.4$ and $\underline{\mu_1}=\underline{\mu}\xi_1/(\xi_1+\xi_2)=1.92$. Hence, $\underline{\mu_2}=\underline{\mu}-\underline{\mu_1}=0.4<\mu_1/3$, and the stochastic system, illustrated in figure~\ref{mr-mu2-positive-stdiag}, tends to follow the stochastic bifurcation diagram shown in figure~\ref{mr-mu2-positive-stdiag}B (corresponding to \ref{mr-eldiag}B). However, the inherent variability in the probabilistic system means that there will also exist events that satisfy $\mu_2>\mu_1/3$ (corresponding to  figure~\ref{mr-eldiag}A). Due to the competition of these two cases, the probability distributions in figure~\ref{mr-mu2-positive-stdiag}B are very diffuse (unlike those in figure~\ref{nh-stdiag}, which are highly correlated around their mean values). Specifically, the distributions around the larger possible stretches are broad for $7<\tau<8$, for example. Thus, it is harder to predict which stretch will be seen because of the higher dispersion of data. One can simulate the case where  $\mu_2-\mu_1/3>0$ in a similar manner (not shown here).

\subsubsection{Negative $\mu_2$}
For negative $\mu_{2}$, it is not enough to ensure that $\mu_2$ is negative, as equilibria only exist in the small finite region  $-\mu_1/5^{5/3}<\mu_2<0$. Using equation \eqref{eq:R12:b}, we set $b=-\mu_1/5^{5/3}$ and define the Beta-distributed random variable
\begin{equation}\label{R1_mu2_negative}
R_1=\frac{\mu_1(1+1/5^{5/3})}{\mu+2\mu_1/5^{5/3}}.
\end{equation}

\begin{figure}[htbp]
	\begin{center}
		A\includegraphics[width=0.45\textwidth]{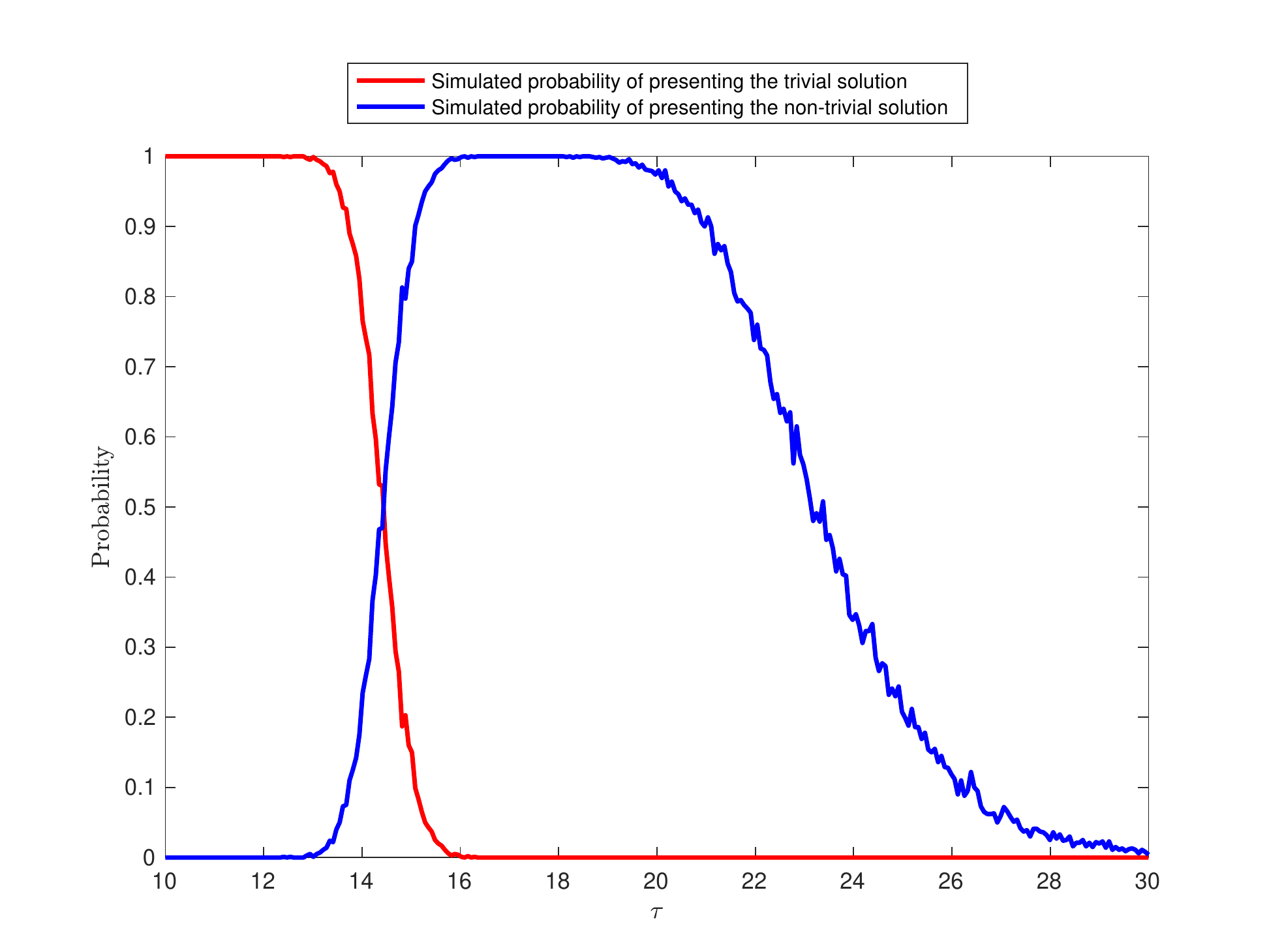}
		B\includegraphics[width=0.45\textwidth]{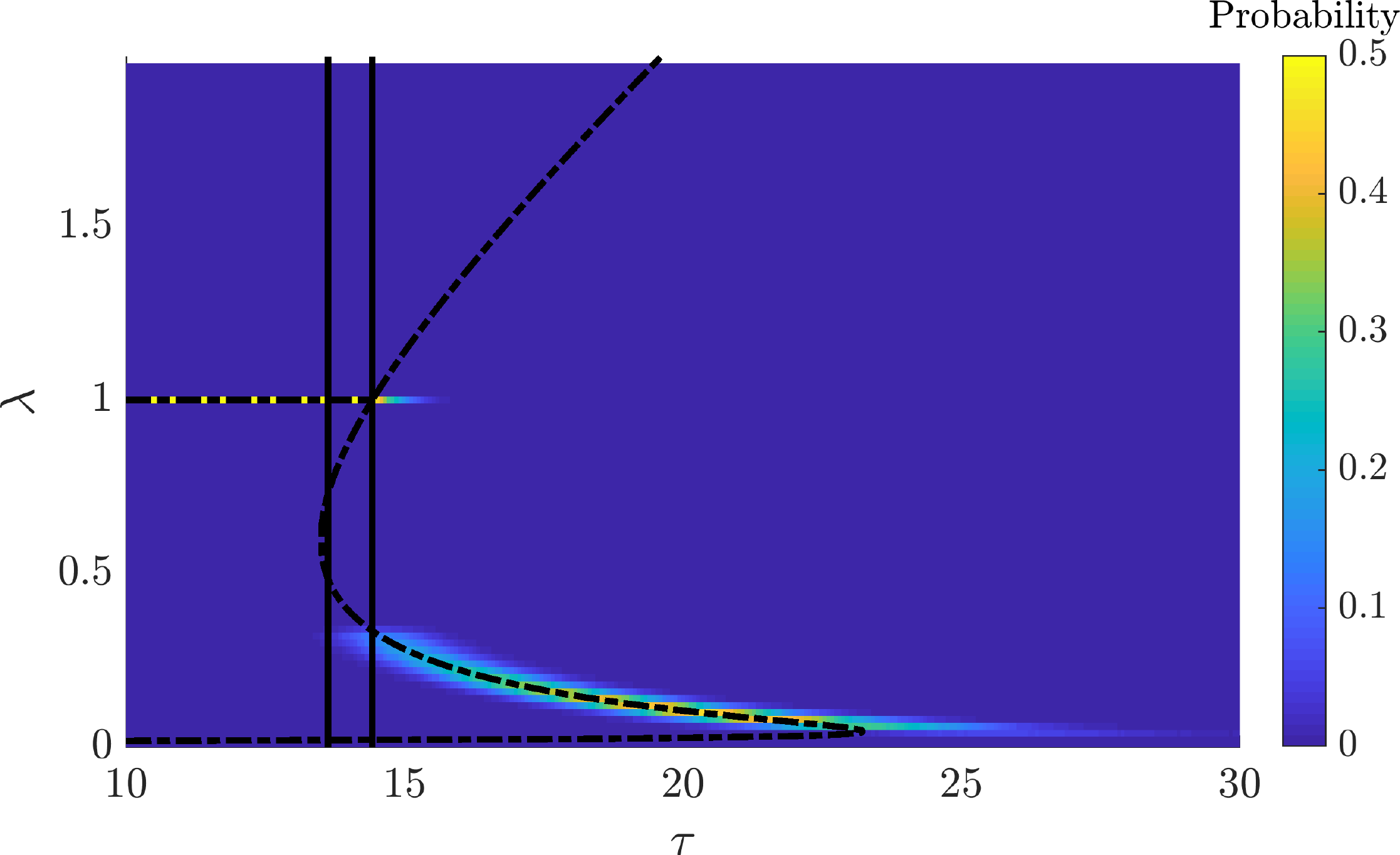}
		\caption{(A) Calculated probability distribution of possible equilibria for a stochastic Mooney-Rivlin cube with $-\mu_{1}/5^{5/3}<\mu_2<0$ and $\rho_1=721$, $\rho_2=0.01$, $\xi_1=10000$, $\xi_2=500$; (B) Stochastic bifurcation diagram with the probability distribution of stretches as function of dead load, $\tau$, following the stable branches of the diagram (see also figure~\ref{mr-eldiag}C).}\label{mr-mu2-negative-stdiag}
	\end{center}
	\begin{center}
		A\includegraphics[width=0.45\textwidth]{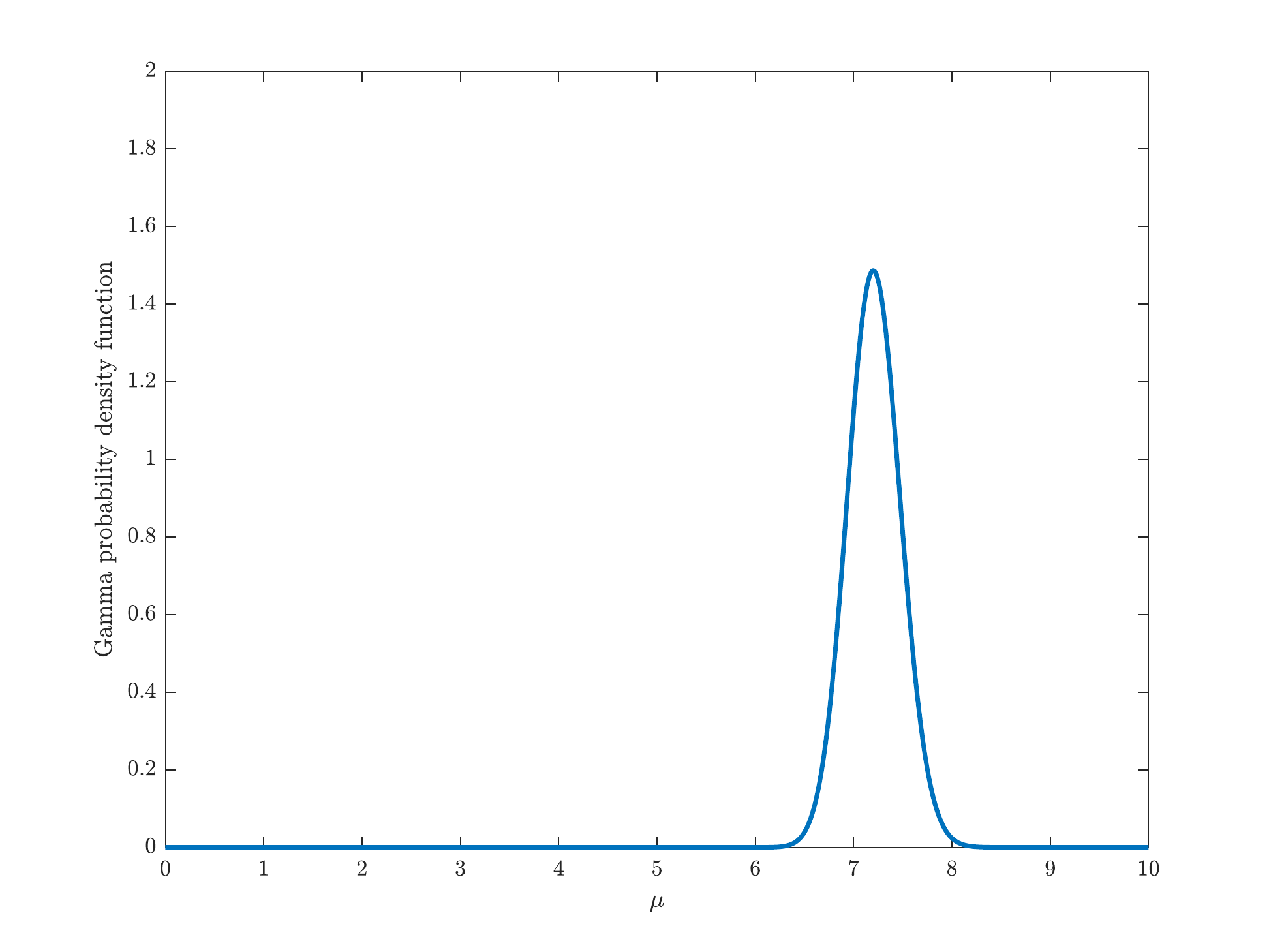}
		B\includegraphics[width=0.45\textwidth]{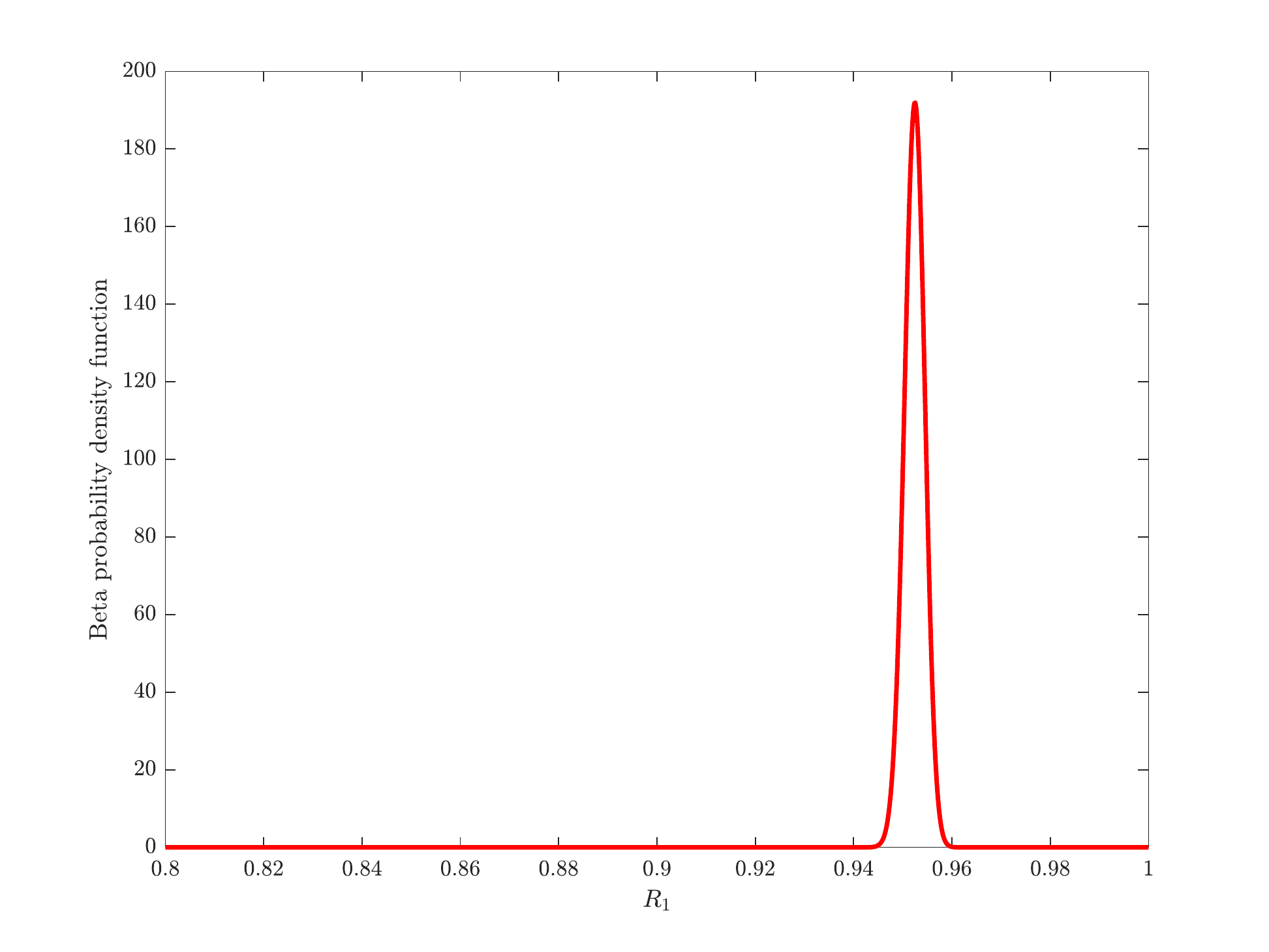}
		\caption{ Assumed (A) Gamma distribution with $\rho_1=721$, $\rho_2=0.01$ for random shear modulus, $\mu=\mu_{1}+\mu_{2}>0$, and (B) Beta distribution with $\xi_1=10000$, $\xi_2=500$ for random variable $R{1}$ given by \eqref{R1_mu2_negative}, in the case of a stochastic Mooney-Rivlin cube with $-\mu_{1}/5^{5/3}<\mu_2<0$.}\label{mr-mu2-negative-gbpdfs}
	\end{center}
\end{figure}

We run similar simulations as in the previous case, noting that we now reproduce $\mu_1$ from equation \eqref{R1_mu2_negative}. Such simulations are illustrated in figure~\ref{mr-mu2-negative-stdiag}, where $\rho_1=721$, $\rho_2=0.01$ and $\xi_1=10000$, $\xi_2=500$ (see figure~\ref{mr-mu2-negative-gbpdfs}). Once again, in figure~\ref{mr-mu2-negative-stdiag}B, we see good correspondence between the stochastic distributions and the deterministic bifurcation diagram in figure~\ref{mr-eldiag}C. Notably, for the stochastic cube population, equilibria exist beyond the maximum value of the deterministic bifurcation curve. Namely, even when $\tau=25$, there is around $10\%$ chance of selecting a cube that presents a non-trivial stretch, whereas this is outside the scope of the deterministic setting.

To summarise, for stochastic Mooney-Rivlin cubes under uniform tensile dead loads, we obtain the probabilities of stable equilibria, given that the model parameters are generated from known probability density functions. In the deterministic elastic case, which is based on mean parameter values, there are single-valued critical loads that strictly separate the cases where either the trivial, reference configuration or a non-trivial stable configuration occurs. By contrast, in the stochastic case, there are probabilistic load intervals, where the trivial and non-trivial states always compete.

\section{Conclusion}\label{sec:conclude}

For a cube of stochastic neo-Hookean or Mooney-Rivlin material subject solely to surface normal dead loads, uniformly distributed in the reference configuration, we have studied possible homogeneous triaxial deformations, and determined which of these deformations are stable. For the deterministic elastic problem, stable non-trivial deformations are possible if the dead loads are sufficiently large. In addition, in the stochastic case, the probabilistic nature of the solution reflects the probability in the constitutive law, and bifurcation and stability can be understood in a probabilistic way. Specifically, by contrast to the deterministic elastic problem, where single-valued critical loads strictly separate the cases where either the trivial configuration or a non-trivial stable configuration occurs, for the stochastic problem, there are probabilistic load intervals, where there is a quantifiable chance for both the trivial and non-trivial states to be found.

This is but one example of the many possible influences of the stochastic hyperelastic model parameters on the predicted nonlinear elastic responses, which cannot be captured by the deterministic approaches. Indeed, the combination of knowledge from elasticity, statistics and probability theories offers a richer set of tools compared to the elastic framework alone, and would logically open the way to further considerations of this type. For example, the limit-point instability of an internally pressurised hyperelastic cylindrical tube, which was also initiated by Rivlin (1949) \cite{Rivlin:1949:VI}, or hyperelastic spherical shell, which was first investigated by Green and Shield (1950) \cite{Green:1950:GS}, and then further studied by Adkins and Rivlin (1952) \cite{Adkins:1952:AR}, are other examples of important elasticity problems which are amenable to the stochastic elasticity framework presented here \cite{Mihai:2018a:MDWG,Mihai:2018b:MDWG}. While these idealised problems are interesting in their own right, they demonstrate also how our stochastic elastic framework captures the variability in the elastic responses of materials under large strains, which are rarely deterministic. 

The Rivlin cube is a rather abstract system that, due to the difficulty of maintaining appropriate boundary conditions during deformations, has no real experimental validation. Yet, its simplicity makes it a paradigm for the understanding of large deformations and stability in elastic materials.  It remains a test case for all new ideas and extensions. As such, Rivlin's intuition in identifying this problem is truly remarkable and its legacy long lasting.  Interestingly, Rivlin had also a strong interest in experiments and obtained himself some beautiful results \cite{Rivlin:1951:VII:RS}. The basic starting point of our approach is that there is more information in data coming from experiments on the elastic properties of materials than just the average values. Due to intrinsic and extrinsic variability, the spread of experimental data also contains important information about the behaviour of materials that should not be discarded. Our stochastic setting provides a clear and systematic method to take that information into account, and will enable us to study more broadly the fascinating nonlinear behaviour of many elastic materials. One can only hope that Rivlin would have enjoyed this \emph{fuzzy twist of chance} on an \emph{old deterministic cube}.

\appendix
\section{Equilibria and stability of homogeneous triaxial stretches under equitriaxial dead loads}\label{sec:append}

In this appendix, we provide a proof of theorem~\ref{th:stability} (see also theorem 2.2 of \cite{Ball:1983:BS}).

\paragraph{Proof: (i)} The fact that there are no pure homogeneous equilibrium configurations other than the reference state when $\tau<0$ follows from the Baker-Ericksen inequalities \eqref{eq:BE}. For the reference configuration, the gradient deformation tensor is the identity tensor, $\textbf{I}=\text{diag}(1,1,1)$. To prove that it is not stable in the sense of definition~\ref{def:stable}, we define the function
\begin{equation}\label{eq:H}
H(\textbf{F})=W(\textbf{F})-\tau\text{tr}(\textbf{F}),
\end{equation}
and take a skew-symmetric matrix $\textbf{A}$ and the exponential mapping
\begin{equation}
e^{\epsilon\textbf{A}}=\sum_{n=0}^{\infty}\frac{\left(\epsilon\textbf{A}\right)^{n}}{n!},
\end{equation}
 where $0<\epsilon\ll 1$. Since $\textbf{A}^{T}=-\textbf{A}$, $\text{tr}(\textbf{A})=0$, it follows that $e^{\epsilon\textbf{A}}$ is proper orthogonal (i.e., $e^{\epsilon\textbf{A}}=\left(e^{\epsilon\textbf{A}}\right)^{T}$ and $\det e^{\epsilon\textbf{A}}=1$).  We consider the following Taylor expansion about the identity tensor in powers of $\varepsilon$
\[
H(e^{\epsilon\textbf{A}})
=W(\textbf{I})-\tau\text{tr}\left(\textbf{I}+\epsilon\textbf{A}+\frac{\epsilon^2}{2}\textbf{A}^2+\mathcal{O}(\epsilon^3)\right)\\
=H(\textbf{I})+\frac{\epsilon^2}{2}\tau\text{tr}\left(\textbf{A}\textbf{A}^{T}\right)+\mathcal{O}(\epsilon^3).
\]
As $\tau<0$ and $\text{tr}\left(\textbf{A}\textbf{A}^{T}\right)>0$, the above identity implies $H(e^{\epsilon\textbf{A}})<H(\textbf{I})+\mathcal{O}(\epsilon^3)$, where $0<\epsilon\ll1$, showing that $\textbf{I}$ is not a local minimum for the total free energy \eqref{eq:energy}. Hence, the reference state is unstable.

\paragraph{(ii)} When $\tau=0$, we consider a deformation $\textbf{y}(\textbf{X})$, such that $\nabla\textbf{y}=\textbf{Q}\text{diag}(\alpha_{1},\alpha_{2},\alpha_{3})\textbf{R}$, where $\textbf{Q}=\textbf{Q}(\textbf{X})$ and $\textbf{R}=\textbf{R}(\textbf{X})$ are orthonormal tensors, satisfying $(\textbf{R}\textbf{Q})_{ii}\leq1$, $i=1,2,3$, and $\alpha_{i}=\alpha_{i}(\textbf{X})$, $i=1,2,3$, are the principal stretches of $\nabla\textbf{y}$, satisfying $\sup\sum_{i=1}^{3}|\alpha_{i}-1|<\epsilon\ll1$ and $\alpha_{1}\alpha_{2}\alpha_{3}=1$. Then, the total free energy \eqref{eq:energy} of the elastic body deformed by $\textbf{y}(\textbf{X})$ is equal to
\[
E(\textbf{y})=\int_{V}\mathcal{W}(\alpha_{1},\alpha_{2},\alpha_{3})dV=E(\textbf{X})+\int_{V}\left[\Psi(\alpha_{1},\alpha_{2},\alpha_{3};\tau)-\Psi(1,1,1;\tau)\right]dV.
\]
From the above expression, we deduce that, if the reference state, $\textbf{X}$, with the gradient tensor $\nabla\textbf{X}=\text{diag}(1,1,1)$, is a strict or non-strict local minimum of $\Psi(\cdot,\tau)$, then $E(\textbf{y})\geq E(\textbf{X})$. Next, taking $\textbf{R}$ proper orthogonal implies $E(\textbf{R})=E(\textbf{X})$, hence the reference state is neutrally stable.

\paragraph{(iii)} When $\tau>0$, we consider a deformation $\textbf{y}(\textbf{X})$ as in (ii).
In this case, the total free energy \eqref{eq:energy} of the elastic body deformed by $\textbf{y}(\textbf{X})$ is equal to
\[
\begin{split}
E(\textbf{y})&=\int_{V}\left[\mathcal{W}(\alpha_{1},\alpha_{2},\alpha_{3})-\tau\text{tr}\left(\textbf{Q}\text{diag}(\alpha_{1},\alpha_{2},\alpha_{3})\textbf{R}\right)\right]dV\\
&=E(\textbf{X})+\int_{V}\left[\Psi(\alpha_{1},\alpha_{2},\alpha_{3};\tau)-\Psi(1,1,1;\tau)\right]dV+\tau\int_{V}\sum_{i=1}^{3}\left[1-(\textbf{R}\textbf{Q})_{ii}\right]\alpha_{i}dV.
\end{split}
\]
From the above expression, we deduce that stability or neutral stability of the reference state, $\textbf{X}$, with the gradient tensor $\nabla\textbf{X}=\textbf{I}$, occurs as follows:

- if $\tau>0$ and the reference state is a strict local minimum of $\Psi(\cdot,\tau)$, as $(\textbf{R}\textbf{Q})_{ii}\leq1$, $i=1,2,3$, then $E(\textbf{y})>E(\textbf{X})$, i.e., the reference state is stable;

- if $\tau>0$ and the reference state is a non-strict local minimum of $\Psi(\cdot,\tau)$, then $E(\textbf{y})\geq E(\textbf{X})$, i.e., the reference state is neutrally stable.
	
\paragraph{(iv)} For a non-trivial homogeneous triaxial stretch $\textbf{x}(\textbf{X})$, with gradient tensor $\nabla\textbf{x}=\text{diag}(\lambda_{1},\lambda_{2},\lambda_{3})$, when $\tau>0$, we consider a deformation $\textbf{y}(\textbf{X})$ as in (iii), and obtain
\[
E(\textbf{y})
=E(\textbf{x})+\int_{V}\left[\Psi(\alpha_{1},\alpha_{2},\alpha_{3};\tau)-\Psi(\lambda_{1},\lambda_{2},\lambda_{3};\tau)\right]dV+\tau\int_{V}\sum_{i=1}^{3}\left[1-(\textbf{R}\textbf{Q})_{ii}\right]\alpha_{i}dV.
\]
In this case, if the triaxial stretch $\textbf{x}(\textbf{X})$ is a (strict or non-strict) local minimum of $\Psi(\cdot,\tau)$, then setting $\nabla\textbf{y}=\textbf{R}\text{diag}(\lambda_{1},\lambda_{2},\lambda_{3})\textbf{R}^{T}$, with $\textbf{R}$ proper orthogonal, implies $E(\textbf{y})=E(\textbf{x})$, hence the triaxial stretch $\textbf{x}(\textbf{X})$ is neutrally stable. This concludes the proof.

\enlargethispage{20pt}

\dataccess{There are no supplementary data associated with this paper.}

\aucontribute{All the authors contributed equally to all aspects of this article and gave final approval for publication.}

\competing{The authors declare that they have no competing interests.}

\funding{The support for Alain Goriely by the Engineering and Physical Sciences Research Council of Great Britain under research grant EP/R020205/1 is gratefully acknowledged.}





\end{document}